\documentclass[a4paper,11pt]{article}
\pdfoutput=1 % if your are submitting a pdflatex (i.e. if you have
\usepackage{jcappub} % for details on the use of the package, please
\usepackage{aas_macros}
\usepackage{hyperref}
\usepackage{graphicx}
\usepackage{subcaption}
\usepackage{float}
\usepackage{color}
\usepackage{cancel}
\usepackage{array}
\usepackage{relsize}
\usepackage{multirow}
\usepackage{orcidlink}
\newcommand{\bea}{\begin{eqnarray}}
	\newcommand{\eea}{\end{eqnarray}}

\title{The pairwise and cross-pairwise y-type polarised kinetic Sunyaev Zeldovich effect from transverse velocity of galaxy clusters}

\author[]{Aritra Kumar Gon\,\orcidlink{0000-0002-0004-9563},}%\note{Corresponding author.}
\author[]{ Rishi Khatri}
\affiliation{Department of Theoretical Physics, Tata Institute of Fundamental Research, \\Mumbai, India}
\emailAdd{aritra.gon@theory.tifr.res.in}
\emailAdd{khatri@theory.tifr.res.in}
\subheader{TIFR/TH/23-19}
\abstract{We develop a new theoretical framework for studying the pairwise and cross-pairwise polarised kinetic Sunyaev Zeldovich (pkSZ) effect arising from the transverse peculiar velocity of galaxy clusters. The pkSZ effect is second order in peculiar velocities and has a spectrum that can be decomposed into y-type and blackbody components, whereas the unpolarised linear kSZ effect has only the blackbody component.  Thus, the detectability of the pkSZ effect depends only on the sensitivity and the number of frequency channels of the survey and not on the other primary and secondary CMB anisotropies. We consider pairing of clusters with other clusters as well as cross-pairing of clusters with galaxies from spectroscopic galaxy surveys. The pairwise pkSZ signal is a function of intra-pair spatial separation. We develop and compare estimators of the pairwise pkSZ effect and study the detectability of the pairwise signal with cluster catalogs consisting of a few hundred thousand clusters expected from surveys such as eROSITA and CMB-S4. We find that cross-pairing clusters with galaxies from a large overlapping spectroscopic survey having a few billion galaxies will enable us to detect the pairwise pkSZ effect with CMB-S4. The pairwise pkSZ effect will thus open up a new window into the large-scale structure of the Universe in the coming decades.}
\begin{document}
	\maketitle
	\flushbottom
	%\newpage
	\section{Introduction \label{sec_intro}}
	During the epoch of reionisation, the radiation from stars and quasars ionises the surrounding neutral hydrogen clouds, producing free electrons.  Post-reionisation, most of the baryons, and therefore free electrons, are present in the regions inside clusters (intra cluster medium - ICM), in the filaments between the galaxy clusters, and around galaxies (circumgalactic medium - CGM). The photons coming from the Cosmic Microwave Background (CMB) interact with these free electrons via Thomson scattering, causing secondary anisotropies in the CMB. These secondary anisotropies are a useful probe of the large-scale structure formation and cosmology in the late time Universe.\\\\
	In this paper, we look at the secondary polarisation of the CMB due to the peculiar velocities of galaxy clusters. Galaxy clusters are the largest collapsed objects in the Universe. Due to gravitational interactions with the large-scale structures around them, the clusters have a peculiar velocity with respect to the CMB rest frame. Thus, the electrons in the clusters also move coherently with the same peculiar velocity and in the rest frame of the electrons the CMB is not isotropic. In particular, there is a quadrupolar anisotropy which is also referred to as the kinematic quadrupole. Thomson scattering of this CMB quadrupole by the free electrons produces polarisation in the CMB. This phenomenon was first predicted by Sunyaev and Zeldovich in 1980 \cite{SZ_80} and is known as the polarised kinetic Sunyaev Zeldovich (pkSZ) effect. Different aspects of the pkSZ effect have been studied previously \cite{1999_Sazonov,2000_Hu,2014_renaux,2016_Pierpaoli,2022JCAP_Hotinli,2022JCAP_Gon}. The previous works mainly focused on the estimation of the kinematic quadrupole and measuring the power spectrum of this polarisation anisotropy signal from reionisation and post-reionisation eras. {\color{black} Since it is a higher-order effect in perturbation theory, the amplitude of the auto-power spectrum is small,  and detecting it would be very challenging. In ref. \cite{2022JCAP_Hotinli}, the authors have analysed the detectability of the pkSZ signal from cross-correlation with reconstructed velocity fields of galaxies. They also developed an quadratic estimator using correlations between CMB polarisation maps and maps of optical depths from some external survey, at small scales. }\\\\
	{\color{black}We propose and study the y-type polarised signal by pairing clusters with other clusters or by pairing clusters with galaxies and develop an estimator for observing the pairwise pkSZ effect. We find that cross pairing clusters with a high signal-to-noise tracer of the peculiar velocity field makes it detectable in upcoming experiments. This is in agreement with similar results obtained for power spectrum in ref. \cite{2022JCAP_Hotinli}.}  After reionisation, the ICM provides the highest optical depth to Thomson scattering and dominates the pkSZ effect. Clusters which are close but well separated from each other tend to move towards each other on average due to their mutual gravitational attraction. As a result, the peculiar velocity vectors of a cluster within a pair will have components that will be aligned. The well-known pairwise linear kinetic SZ effect deals with the change in intensity of the CMB radiation due to the radial peculiar velocity of the clusters. The intensity of the scattered radiation is blueshifted (redshifted) for the cluster in the pair further from (closer to) us along the line of sight. Therefore, if we subtract the intensity from two different clusters and average over many cluster pairs, we get a non-zero signal which is proportional to the mean pairwise relative velocity. The mean pairwise velocity, which was initially introduced in the context of BBGKY theory \cite{1980_peebles_lss,1977ApJ_Davis,1994ApJ_PerturbativeGrowth}, is a function of separation between the clusters. The pairwise kSZ effect has been studied by several authors in the past \cite{1999ApJ_DynamicalEstimator,2012_Hand,2015ApJ_DarkEnergy,2015PRD_massiveneutrinos,2016PRL_ProjectedFields:ANovelProbe}. More recently, it also has been detected experimentally \cite{2016_DESYear1andSPT,2018_PlanckandBOSSdata,2021_PRD_AtacamaCosmologyTelescope,2022MNRAS_DESIgalaxyclustersandPlanck,2023_SPT_3GandDESCollaborations}. Conceptually, our estimator is similar to the pairwise kinetic SZ  effect, nonetheless, there are some important differences.\\\\
	For the pkSZ effect, we are interested in polarisation, which is a spin-2 field and arises from the transverse component of the peculiar velocity in contrast to the radial velocity for the kSZ effect. The polarisation direction of the scattered photons is always perpendicular to the transverse velocity direction. Moreover, it depends only on the square of the transverse velocity and remains unchanged if we flip the direction of the transverse velocity vector. Thus, in our case, instead of subtracting we add the polarisation from two clusters and perform the averaging over all cluster pairs at a given separation. This procedure defines our pairwise estimator. We show that the resulting pairwise polarisation signal depends on the separation between the clusters and cosmological parameters such as the Hubble parameter, growth factor, growth rate, and the linear matter power spectrum. It also depends on astrophysical parameters like the mean optical depth of the cluster as well as on the halo bias factors.\\\\
	The reason the pairwise estimator works is that, for a cluster in a pair, the second cluster acts as an indicator of the direction in which the first cluster is moving. Ideally, we would like to rotate the clusters so that the velocities are aligned and then stack them, so that the signal adds up coherently. However, we do not know the direction of motion for each cluster and instead have to use the correlation of the peculiar velocity of a cluster with the nearby clusters. In general, we can pair up a cluster with any other indicator of the large-scale gravitational potential around the cluster responsible for its peculiar velocity. We show that with future galaxy surveys, with billions of galaxies, cross-pairing clusters with galaxies would be a more promising observable. Another interesting possibility will be cross-correlation with the 21-cm surveys.\\\\
	We calculate the detectability and signal-to-noise ratio for the pkSZ signal from future CMB experiments.  In observations, we average over different pairs of clusters which are at a given separation. In general, different cluster pairs will have different mean masses or optical depths and redshifts. Thus, the cosmological factors associated with the polarisation signal will also be different for different pairs. Moreover, the separation vector will be oriented differently with respect to the mean line of sight direction. Averaging over cluster pairs, therefore, includes averaging over all of these variables. \\\\
	We estimate the noise associated with the measurement of the pairwise pkSZ effect for a survey with typical values of sensitivity and angular size of the beam expected for future CMB experiments such as  CMB-S4  \cite{2016_cmbs4} wide survey and CMB-HD \cite{sehgal2019cmb}. We also present our results as the number of cluster pairs required in a separation bin to measure the pairwise signal at a signal-to-noise ratio of 1. We further analyse the situation in which we pair clusters from the CMB experiments with galaxies from large surveys like LSST \cite{2009_LSST} having complete sky overlap. We show that with spectroscopic galaxy catalogs, the pairwise pkSZ effect could be detected at high significance. The pairwise pkSZ effect thus opens up a new pathway for cosmology. \\\\
	We assume a flat $\mathrm{\Lambda CDM}$ universe with baryon and matter density parameters $\mathrm{\Omega_b = 0.0490}$ and $\mathrm{\Omega_m = 0.3111}$, Hubble constant, $\mathrm{H_0} = 100\,\mathrm{h}$ $\mathrm{km\,s^{-1} Mpc^{-1}}$ with $\mathrm{h = 0.6766}$, spectral index of primordial curvature perturbations $\mathrm {n_s = 0.9665}$, its amplitude $\mathrm {log(A_s)} =-8.678 $ and helium mass fraction $\mathrm{X_{He} = 0.24}$ \cite{aghanim2020planck}. We use publicly accessible codes CAMB \cite{lewis2000efficient} and COLOSSUS \cite{diemer2018colossus} for our numerical analysis. We use units with the speed of light in vacuum $c=1$.
	\section{Pairwise pkSZ effect from galaxy clusters \label{sec_main_derivation}}	 
	The linear polarisation of CMB radiation from the Thomson scattering can be characterised by the pair of Stokes parameters Q and U.  We define $\left\{\mathcal{Q},\mathcal{U}\right\}\equiv\left\{\mathrm{\frac{Q}{I},\frac{U}{I}}\right\}$, where I is the average background intensity. Furthermore, it is easier to use the combination $ P_{\pm}\equiv\left(\mathcal{Q}\pm i\mathcal{U}\right)$ whose modulus gives the amplitude of polarisation relative to the background intensity. Since $P_{+}$ and $P_{-}$ are related by complex conjugation, henceforth we only use $P_{+}$ for our calculations.  In the non-relativistic limit, the amplitude of the polarisation from a point $\mathbf{ r}$ inside the ICM of a galaxy cluster is given by \cite{2022JCAP_Gon},
	\begin{align}
		\label{intro_eqn_1}
		P_{+}(\mathbf{ \hat{n}})=-\frac{\sqrt{6}}{10}\sigma_{\mathrm{T}}\int d\chi\,e^{-\tau(\chi)}\, n_e({\mathbf{r}})\,a(\chi)\sum^{2}_{\lambda=-2}\,_2Y_{2\lambda}(\mathbf{\hat{n}})\;\sum_{i,j=1}^{3}v_{i}({\mathbf{x}})v_{j}({\mathbf{x}})\int \hat{n}'_i \hat{n}'_j \,Y_{2\lambda}^{*}(\mathbf{\hat{n}'})\,d\Omega_\mathbf{\hat{n}'},
	\end{align}	
	where $\sigma_{\mathrm{T}}$ is the Thomson scattering cross section, $\,_2Y_{2\lambda}$ are the spin-2 spherical harmonics, $\mathbf{\hat{n}}$ is the line of sight direction, $\mathbf{\hat{n}'}$ is the direction of incoming photons, and $\mathbf{x}$ is the position vector to the center of the cluster. The comoving distance $\chi$ is related to the conformal time $\eta$ as $\chi =\eta_0-\eta$, where $\eta_0$ is the conformal time today. The integral over $\chi $ is along a line of sight through the cluster. The frequency spectrum of the polarised intensity is a combination of the blackbody radiation and y-type spectrum. There are a few points to note here. Each cluster has an electron number density profile  $n_e({\mathbf{r}})$ in the ICM which depends on the mass of the cluster and at the point $\mathbf{ r}$ inside the cluster. The line of sight integration is through the galaxy cluster. The velocity $\mathbf{v}(\mathbf{x})$ is the peculiar velocity of the cluster and  $i$ and $j$ are indices denoting the components of the velocity vector. 
	%We use linear theory to produce this velocity field, i.e. we assume that the clusters are moving with a velocity which is equal to the dark matter velocity field at the centre of the cluster. There is an underlying course graining which is happening because of this approximation. This is a very standard assumption as clusters are very large and so linear theory approximations should be okay. Secondly, velocities are sourced by the gravity of all matter present and not just the baryonic matter,  so the bias factor associated with velocity fields for a cluster is negligible.
	We can rewrite Eq.(\ref{intro_eqn_1}) in a simpler form, if we rotate the coordinates such that line of sight direction coincides with the z axis. Furthermore, we can approximate $e^{-\tau(\chi)}\approx1$ in Eq.(\ref{intro_eqn_1}). Doing so, we get \cite{2022JCAP_Gon},
	\begin{align}
		\label{intro_eqn_2}
		P_{+}{\left(\hat{\mathbf{n}}\equiv\hat{\mathbf{z}}\right)}=-\frac{\sigma_{\mathrm{T}}}{10}\int d\chi\,\,n_\mathrm{e}(\mathbf{ r})a(\chi)v_t^2(\mathbf{x})\;e^{-2i\phi},
	\end{align}	
	\begin{figure}%[h!]
	\centering
	%\hspace{3.5cm}
	\begin{subfigure}{0.83\textwidth}
		%	\vfill
		\centering
		\includegraphics[width=0.86\linewidth]{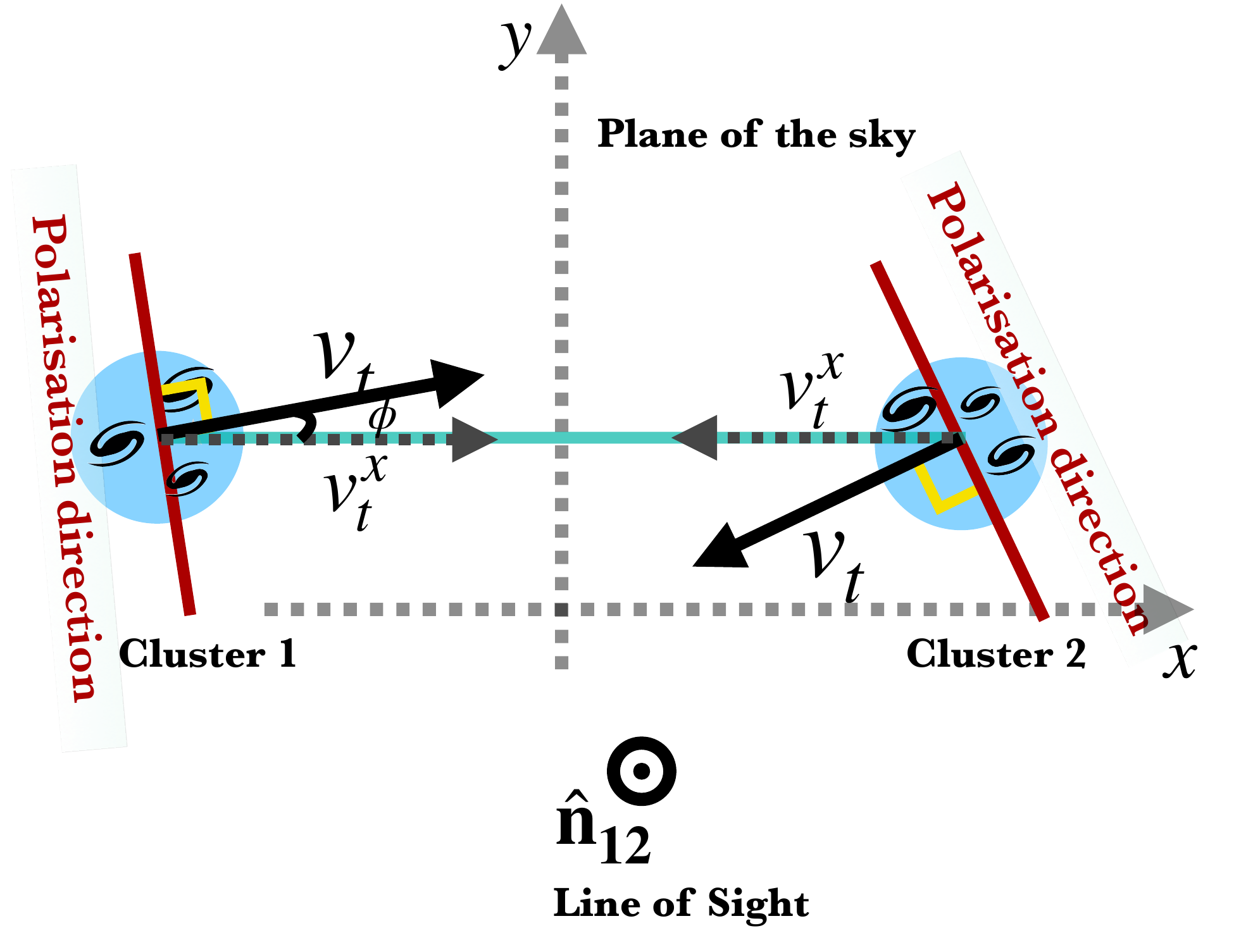}
	\end{subfigure}
	\caption{The plane defined by the x-y axes is the plane of the sky. The two clusters are the two blue spheres which are moving towards each other under their mutual gravitational attraction. The projection of the two clusters on the plane of the sky has been shown here. The transverse velocity is given by $v_t$. We see that if the transverse velocities are almost aligned, there will be a component of polarisation which adds up coherently when we average over many cluster pairs.}
	\label{fig:cluster_pair}
\end{figure}\begin{figure}%[H]
		\centering
		%\hspace{3.5cm}
		\begin{subfigure}{1.0\textwidth}
			%	\vfill
			\centering
			\includegraphics[width=0.98\linewidth]{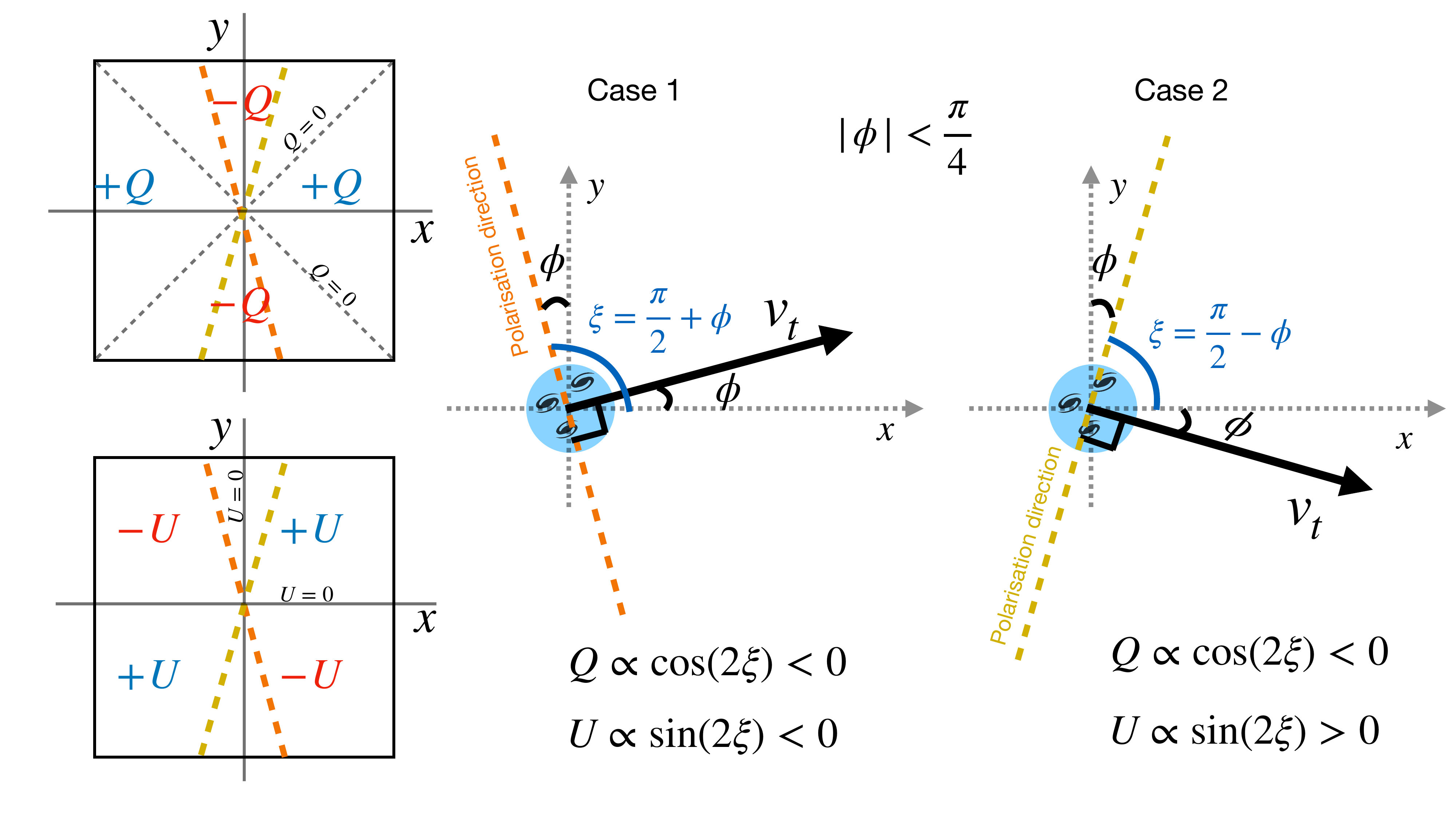}
		\end{subfigure}
		\caption{The x-y plane is the plane of the sky. The transverse velocity vector makes and angle $\phi$ with the x-axis and since the clusters are attracted to each other, we can assume $|\phi|<\frac{\pi}{4}$ for close clusters. In this choice of our coordinates, the $\mathcal{Q}$ parameter is always negative, while depending on the direction of $\mathbf{v_t}$, $\mathcal{U}$ parameter can be of either sign.}
		\label{fig:pol_perp}
	\end{figure}where $\phi$  is the angle between the transverse (to the line of sight direction) velocity vector $\mathbf{v_t}$  of the electron and x-axis of our chosen coordinate system on the tangent plane (plane of the sky) as shown in figure \ref{fig:cluster_pair}. We see from Eq.(\ref{intro_eqn_2}) that the polarisation direction is always perpendicular to $\mathbf{v_t}$.  This can be easily understood if we rotate our coordinate system such that $\mathbf{v_t}$ aligns with the x-axis i.e. $\phi=0$, then $\mathcal{U}=0$ and $\mathcal{Q}<0$. Thus, the linear polarisation is along the y-axis with this choice of coordinates. If a pair of clusters are close, their peculiar velocity will have a greater probability of being aligned due to their mutual gravitational attraction. Thus, the transverse component of the peculiar velocities on the plane of the sky will also tend to be aligned. 
For a pair of clusters, we choose our coordinates such that the x-axis aligns with the projection of the spatial separation vector between the pair on the plane of the sky as shown in figure \ref{fig:cluster_pair}. The misalignment of the $\mathbf{v_t}$ of a cluster with respect to the x-axis (angle $\phi$) due to other structures around the cluster will fluctuate randomly about the x-axis from pair to pair as shown in figure \ref{fig:cluster_pair}. The net polarisation from each cluster in the pair can be described by their respective $\mathcal{Q}$ and $\mathcal{U}$ parameters. Thus, in our coordinate system, if we add up the polarisation from two clusters and average over many such pairs which are at a fixed separation, the $\mathcal{Q}$ parameter will always have the same negative sign and will add up coherently, while the $\mathcal{U}$ parameter can have either sign and will average to 0 when averaging over many cluster pairs as shown in figure \ref{fig:pol_perp}. Because of this coherent addition of the $\mathcal{Q}$ parameter, we should get a net non-zero polarisation. In a general coordinate system but with separation vectors of all pairs still aligned before stacking, both $\mathcal{Q}$ and $\mathcal{U}$ parameters will be non-zero, but the modulus of $\mathcal{Q} + i\mathcal{U}$ remains invariant. The reason we add polarisation from the cluster pairs in the pkSZ case, instead of subtracting as we do with intensities in pairwise kSZ, is that the polarisation direction remains invariant under a rotation by $\pi$. It does not depend on whether the cluster is moving in the $+x$ or $-x$ direction unlike in kSZ where the intensity either redshifts or blueshifts depending on the position of the cluster.
\subsection{Estimator for the pairwise pkSZ effect.  \label{pksz_estimator}}
We can formally define the estimator for the pairwise pkSZ effect as,
\begin{align}
	\hat{P}_{\mathrm{pairwise}}(x)=\sum_{\mathrm{ i}}w_i\,(P_{i1+}+P_{i2+})\Big|_{\mathrm{separation\,along\,x-axis}},
\end{align}
where the sum is over all distinct cluster pairs. $P_{i1+}$ and $P_{i2+}$ denote the polarisation measured from two different clusters in a pair, separated by a given distance x,  in a coordinate system in which the x-axis is along the separation vector $\mathbf{ x}$. Considering isotropy and homogeneity of space at large scales, we expect that the estimator should only depend on the magnitude of separation between clusters. The normalised weights are denoted by $w_i$ which we want to optimize to maximize the signal-to-noise ratio.  The most simple case is that of uniform weighting, $w_i=1/\sum_{\mathrm{N_{pairs}}}$, where $\mathrm{N_{pairs}}$ is the total number of cluster pairs. This weighting is however not optimal and we will consider optimizing the weights using heuristic arguments inspired by linear kSZ estimator in section \ref{s_n_weighted}. The expectation value of the estimator is given by,
\begin{align}
	\Big\langle\hat{P}_{\mathrm{pairwise}}\Big\rangle_2=\sum_{\mathrm{i}}w_i	\Big\langle(P_{i+}+P_{i+})\Big\rangle_2,
\end{align}
where the subscript 2 on the angular brackets $\langle \cdots \rangle_2 $ indicates that the ensemble average is over the two-particle (or two-cluster) distribution function. We shall now lay out the framework for estimating the ensemble average over the two-particle distribution function analytically.
	\subsection{Theoretical formalism for the pairwise pkSZ signal  \label{Theory_derivation}}
	Considering two points ${\mathbf{r_1}}$ and ${\mathbf{r_2}}$ (in comoving coordinates) in the ICM of the two clusters, the ensemble average can be formally expressed as,
	\begin{align}
		\label{ensemble_1}
		&P_{\mathrm{pairwise}}(\mathbf{x_1},\mathbf{x_2},\mathbf{\hat{n}_1},\mathbf{\hat{n}_2}|m,\chi)\equiv\Big\langle P_{+}\left(\mathbf{\hat{n}_1}\right)+P_{+}\left(\mathbf{\hat{n}_2}\right)\Big\rangle_2=\nonumber\\
		&-\frac{\sqrt{6}}{10}\sigma_{\mathrm{T}}\Bigg[\int d\chi_1\,n_{e1}({\mathbf{r_1}})\,a(\chi_1)\sum_{\lambda,i,j}\,_2Y_{2\lambda}(\mathbf{\hat{n}_1})\Big\langle v_{1i}({\mathbf{x_1}})v_{1j}({\mathbf{x_1}})\Big\rangle_2\int \hat{n}'_i \hat{n}'_j Y_{2\lambda}^{*}(\mathbf{\hat{n}'})d\Omega_\mathbf{\hat{n}'}\:+\nonumber\\
		&\hspace{0.8cm}\int d\chi_2\,n_{e2}({\mathbf{r_2}})\,a(\chi_2)\sum_{\lambda,i,j}\,_2Y_{2\lambda}(\mathbf{\hat{n}_2})\Big\langle v_{2i}({\mathbf{x_2}})v_{2j}({\mathbf{x_2}})\Big\rangle_2\int \hat{n}'_i \hat{n}'_j Y_{2\lambda}^{*}(\mathbf{\hat{n}'})d\Omega_\mathbf{\hat{n}'}\Bigg],
	\end{align}				
	The subscripts $1$ and $2$ in all the variables denote the corresponding quantities in the two different clusters. The line of sight directions are given by $\mathbf{\hat{n}_1}$ and $\mathbf{\hat{n}_2}$ and the vectors ${\mathbf{x_1}}$ and ${\mathbf{x_2}}$ denote the position vectors of the centre of the two clusters. Henceforth, we will suppress ${\mathbf{x_1}}$ and ${\mathbf{x_2}}$ arguments in the velocities for brevity. To proceed further, we make a few assumptions. We assume that the two clusters are situated at some average redshift z, but are separated by a distance given by $x=|\mathbf{ x_1}-\mathbf{ x_2}|$. We also assume that $\mathbf{\hat{n}_1}\simeq \mathbf{\hat{n}_2}\simeq(\mathbf{\hat{n}_1}+ \mathbf{\hat{n}_2})/2=\mathbf{\hat{n}_{12}}$. Finally, we also consider the two clusters to have the same mass given by the average mass of the two clusters and thus they have the same effective optical depth $\tau_{\mathrm{eff}}$ (defined in section \ref{effective-opt_depth}). All these approximations are very similar to the ones used in pairwise linear kSZ effect estimation \cite{1980_peebles_lss,1977ApJ_Davis}. Using these approximations we can rewrite Eq.(\ref{ensemble_1}) as,
	\begin{align}
		\label{main_b}
		&P_{\mathrm{pairwise}}(\mathbf{x_1},\mathbf{x_2},\mathbf{\hat{n}_{12}}|m,\chi)=-\frac{\sqrt{6}}{10}\: \tau_{\mathrm{eff}}\sum_{\lambda,i,j}\,_2Y_{2\lambda}(\mathbf{\hat{n}_{12}})\Big\langle v_{1i}v_{1j}+v_{2i}v_{2j}\Big\rangle_2\int \hat{n}'_i \hat{n}'_j Y_{2\lambda}^{*}(\mathbf{\hat{n}'})d\Omega_\mathbf{\hat{n}'}.
	\end{align}	
	The expectation value of the velocities denoted by $\langle\cdots \rangle_2$ is defined as \cite{1980_peebles_lss,1992ApJ_Fry},
	\begin{align}
		\label{ensemble_defn_1}
		\Big\langle v_{1i}v_{1j}+v_{2i}v_{2j}\Big\rangle_2(\mathbf{x_1},\mathbf{x_2}|m,\chi)\equiv\frac{\int\int\left(v_{1i}v_{1j}+v_{2i}v_{2j}\right)f_2(\mathbf{x_1},\mathbf{x_2},\mathbf{v_1},\mathbf{v_2}|\chi,m)d^3\mathbf{v_1}d^3\mathbf{v_2}}{\int\int f_2(\mathbf{x_1},\mathbf{x_2},\mathbf{v_1},\mathbf{v_2}|\chi,m)d^3\mathbf{v_1}d^3\mathbf{v_2}},
	\end{align}
	where $f_2(\mathbf{x_1},\mathbf{x_2},\mathbf{v_1},\mathbf{v_2}|\chi,m)$ is the two particle distribution function of the halos in the position-velocity phase space for a given mass m at a comoving distance $\chi$. Note that, we can replace momentum  with velocity as one of the phase space variables in the non-relativistic limit. The two particle phase space distribution of halos can be related to the local joint probability distribution of halo number density and velocity fields  \cite{2004ApJ_Ma_CosmologicalKineticTheory,1992ApJ_Fry}, $p(n_1,n_2,\mathbf{v_1},\mathbf{v_2}|\chi,m)$  as,
	\begin{align}
		\label{map_1}
		f_2(\mathbf{x_1},\mathbf{x_2},\mathbf{v_1},\mathbf{v_2}|\chi,m)=\int\int n_1(\mathbf{x_1})n_2(\mathbf{x_2})\;p(n_1,n_2,\mathbf{v_1},\mathbf{v_2}|\chi,m) \color{black}dn_1dn_2 ,
	\end{align}
	where $n(\mathbf{x}|m,\chi)$ denotes the local number density of halos with mass m at comoving distance $\chi$, centred at $\mathbf{x}$. The local number density $n(\mathbf{x}|m,\chi)$ can be expressed as,
	\begin{align}
		\label{halo_overdensity_1}
		n(\mathbf{x}|m,\chi)=\bar{n}(m,\chi)(1+\delta^h(\mathbf{x},m,\chi)),
	\end{align}
	where $\bar{n}(m,\chi)$ is the halo mass function \cite{2008_Tinker} and $\delta^h$ is the local halo overdensity field. Using Eq.(\ref{map_1}) and Eq.(\ref{halo_overdensity_1}) in Eq.(\ref{ensemble_defn_1}), we can write the numerator as,
	\begin{align}
		&\int\int\int\int\left(v_{1i}v_{1j}+v_{2i}v_{2j}\right)n_1(\mathbf{x_1})n_2(\mathbf{x_2})\,p(n_1,n_2,\mathbf{v_1},\mathbf{v_2},|\chi,m)\,\color{black}dn_1dn_2\color{black}\,d^3\mathbf{v_1}d^3\mathbf{v_2}=	\nonumber\\
		&\hspace{5cm}\bar{n}^2\Big\langle\left(v_{1i}v_{1j}+v_{2i}v_{2j}\right) (1+\delta_1^h)(1+\delta_2^h)\Big\rangle,
	\end{align}
	where the angular brackets indicates the usual ensemble average over the density and velocity fluctuations. Similarly, the denominator in Eq.(\ref{ensemble_defn_1}) can be written as,
	\begin{align}
		&\int\int\int\int n_1(\mathbf{x_1})n_2(\mathbf{x_2})\,p(n_1,n_2,\mathbf{v_1},\mathbf{v_2},t|m)\,\color{black}dn_1dn_2\color{black} \,d^3\mathbf{v_1}d^3\mathbf{v_2}=	\bar{n}^2\Big\langle (1+\delta_1^h)(1+\delta_2^h)\Big\rangle.
	\end{align}
	Thus, putting the numerator and denominator together we finally get, 
	\begin{align}
		\label{ensemble_defn_2}
		\Big\langle v_{1i}v_{1j}+v_{2i}v_{2j}\Big\rangle_2=\frac{\Big\langle (v_{1i}v_{1j}+v_{2i}v_{2j})(1+\delta^h_1)(1+\delta^h_2)\Big\rangle}{\Big\langle (1+\delta^h_1)(1+\delta^h_2)\Big\rangle}.
	\end{align}
	The local halo overdensity is related to the local dark matter overdensity as, \cite{2010PRD_Largescalevelocities_Fabian,2002PhR_Cooray_Sheth,1997MNRAS_Mo_White}
	\begin{align}
		\label{halo_dm_reln}
		\delta^h(\mathbf{x},\chi;m,z)=\sum_{p}\frac{b_p(m,z)}{p!}\delta^p(\mathbf{x},\chi),
	\end{align}
	where $\delta(\mathbf{x},\chi)$ is the dark matter density field and $b_p(m,z)$ are the halo bias parameters. We assume the halo velocities to be unbiased tracers of the matter velocity field \cite{2018ApJ_velbias}. The dark matter overdensity field $\delta$, can be expanded in perturbation theory as, \cite{2002PhR_Bernardeau,1994ApJ_Bhuvnesh}
	\begin{align}
		\label{dm_pert_exp}
		\delta(\mathbf{x},\chi)=\sum_{q}\delta^{(q)}(\mathbf{x},\chi),
	\end{align}
	where $\delta^{(1)}$ is linear density field and $\delta^{(q)}$  is $q^{th}$ order density field. We can now use  Eq.(\ref{halo_dm_reln}) and Eq.(\ref{dm_pert_exp}) in Eq.(\ref{ensemble_defn_2}) to calculate the ensemble average. Let us first look at the numerator. We do a perturbative expansion of $\delta^h_1$ and $\delta^h_2$, collecting all the terms up to $4^{th}$ order in perturbation theory. We note that the $v_iv_j$ term coming from one of the clusters is already at second order and we want to correlate it with the density perturbation of the second cluster. Since we assume that the initial perturbations are Gaussian, all the odd N-point functions will be zero. Therefore, the first non-zero contributions come from the  $4^{th}$ order terms with velocity fields at linear order. The numerator of Eq.(\ref{ensemble_defn_2}) at leading order is,
	\begin{align}
		\label{contri_terms}
		&\Big\langle (v_{1i}v_{1j}+v_{2i}v_{2j})(1+\delta^h_1)(1+\delta^h_2)\Big\rangle = b_1\Big\langle v_{1j}^{(1)}v_{1j}^{(1)}\delta_2^{(2)}\Big\rangle\:+b_1 \Big \langle v_{2i}^{(1)}v_{2j}^{(1)}\delta_1^{(2)}\Big\rangle+\nonumber\\
		&\hspace{4.cm} \frac{b_2}{2} \Big \langle v_{1i}^{(1)}v_{1j}^{(1)}\left(\delta_2^{(1)}\right)^2\Big\rangle+ \frac{b_2}{2} \Big \langle v_{2i}^{(1)}v_{2j}^{(1)}\left(\delta_1^{(1)}\right)^2\Big\rangle\;+ \mathrm{higher\;order\;terms}.
	\end{align}
	To proceed further we move to Fourier space. In Fourier space, the density fields at first and second order and  the velocity fields at first order in perturbation theory are given by \cite{2002PhR_Bernardeau,1994ApJ_Bhuvnesh},
	\begin{align}
		\label{delta1}
		&\delta^{(1)}(\mathbf{x},\chi)=D(\chi)\int\frac{d^3\mathbf{k}}{(2\pi)^3}\exp\left(i \mathbf{k}\cdot\mathbf{x}\right)\delta^{(1)}(\mathbf{k}),
	\end{align}
	\begin{align}
			\label{delta2}
		&\delta^{(2)}(\mathbf{x},\chi)=D^2(\chi)\int\frac{d^3\mathbf{k}}{(2\pi)^3}\exp\left(i \mathbf{k}\cdot\mathbf{x}\right) \int\frac{d^3\mathbf{q_1}d^3\mathbf{q_2}}{(2\pi)^3}\:\delta^3\left(\mathbf{q_1}+\mathbf{q_2}-\mathbf{k}\right)\nonumber\\
		&\hspace{3cm}
		\left[\frac{5}{7}+\frac{2}{7}\frac{(\mathbf{q_1}\cdot\mathbf{q_2})^2}{q_1^2\,q_2^2}+\frac{(\mathbf{q_1}\cdot\mathbf{q_2})}{2}\left(\frac{1}{q_1^2}+\frac{1}{q_2^2}\right)\right]\delta^{(1)}(\mathbf{q_1})\delta^{(1)}(\mathbf{q_2}),
	\end{align}
	\begin{align}
			\label{vel1}
		&\mathbf{v}^{(1)}(\mathbf{x},\chi)=D(\chi)\,[afH](\chi)\int\frac{d^3\mathbf{k}}{(2\pi)^3}\exp\left(i \mathbf{k}\cdot\mathbf{x}\right) \frac{i\mathbf{k}}{k^2}\delta^{(1)}(\mathbf{k}),
	\end{align}
		where $\delta^{(1)}(\mathbf{k})=\mathcal{R}(\mathbf{k})T(k)$, where $\mathcal{R}(\mathbf{k})$ is the primordial curvature perturbation and $T(k)$  is the transfer function. $D(\chi)$ is the growth factor normalised as D(0)=1. The growth rate $f={d\ln D}/{d \ln a}$. 
	In $\Lambda$CDM cosmology we have $f=\Omega^{0.55}_m(\chi)$. We can represent the overdensity and peculiar velocity in a separable equation in $\chi$ and $\mathbf{k}$, and obtain analytic solutions given in  Eq.(\ref{delta1}),  Eq.(\ref{delta2}), and  Eq.(\ref{vel1}) if we approximate $f=\Omega^{0.55}_m(\chi)\approx\Omega^{0.5}_m(\chi)$ \cite{2002PhR_Bernardeau}.\\\\
The second and the fourth terms in Eq.(\ref{contri_terms}) are identical to the first and the third terms respectively if we make the substitution, $\mathbf{k_1}\rightarrow-\mathbf{k_1}$ and $\mathbf{k_2}\rightarrow-\mathbf{k_2}$. We, therefore, only have to evaluate the first term and the third term. The details of the calculations are given in Appendix  \ref{App:Derivation_details}. Similarly, we can evaluate the denominator, in Eq.(\ref{ensemble_defn_2}). In the denominator, the 2-point correlation function is non-zero and gives the leading order contribution. Therefore, we do not need to go to higher-order terms which are subdominant. After a lengthy calculation and performing the angular integrals, details of which are given in Appendix \ref{App:Derivation_details}, we obtain
	\begin{align}
		\label{final_result_1}
		&P_{\mathrm{pairwise}}(\mathbf{x},\mathbf{\hat{n}_{12}}|m,\chi)=\Bigg[\frac{\sqrt{\pi}}{10\pi^5}\,D^4(Hfa)^2\,
		\tau_{\mathrm{eff}}\;Y_{2-2}(\mathbf{\hat{x}};\mathbf{\hat{n}_{12}})\sum_{L_1,L_2}\sum^{2}_{q=0}\sum_{l}i^{(L_1+L_2)}\,(-1)^{(L_1+1)}\nonumber\\
		&(2L_1+1)(2L_2+1)(2l+1)\frac{q!}{(q-l)!!(q+l+1)!!}
		\left(\begin{array}{ccc}
			1& L_1 & l\\ 
			0& 0 & 0
		\end{array}\right)
		\left(\begin{array}{ccc}
			1& L_2& l\\ 
			0& 0 & 0
		\end{array}\right)
		\left(\begin{array}{ccc}
			L_1& L_2 & 2\\ 
			0& 0 & 0
		\end{array}\right)
		\left\{\begin{array}{ccc}
			2& L_2 & L_1\\ 
			l& 1& 1
		\end{array}\right\}\nonumber\\
		&\int dk_1dk_2k_1^2k_2^2\; G_{q}(k_1,k_2,b_1,b_2)j_{L_1}(k_1x)\,j_{L_2}(k_2x)P(k_1)P(k_2)\Bigg]\Bigg[1+\frac{D^2b_1^2}{2\pi^2}\int dkk^2j_0(kx)P(k)\Bigg]^{-1},
	\end{align}
where $P(k)=P_{\mathcal{R}}(k)\,T^2(k)$ and $P_{\mathcal{R}}(k)$ is the power spectrum of the initial curvature perturbation. The summation index $l$ assumes the values $l=0, 2, \cdots, q-2, q$ if q is even, and $l = 1, 3, \cdots,
	q-2, q$ if q is odd,  and
	$\left(\begin{array}{ccc}
		l_1& l_2 & l_3\\ 
		m_1& m_2& m_3
	\end{array}\right)$  and 
	$
	\left\{\begin{array}{ccc}
		l_1& l_2 & l_3\\ 
		m_1& m_2& m_3
	\end{array}\right\}$ are the Wigner-3j symbol and 6j symbol respectively, following the conventions from  \cite{varshalovich1988quantum}. We can rewrite Eq.(\ref{final_result_1}) as a function of the redshift (comoving distance) and mass dependent factors only,

	\begin{align}
		\label{final_result_Mz_form}
	 &P_{\mathrm{pairwise}}(\mathbf{x},\mathbf{\hat{n}_{12}}|m,\chi)=A\left[(D^2Hfa)^2\right]\hspace{-0.15cm}(\chi)\,\tau_{\mathrm{eff}}(m,\chi)\nonumber\\
		&\hspace{5.7cm} \left[\frac{b_1(m,\chi)\,C_1(x)+b_2(m,\chi)\,C_2(x)}{1+D^2(\chi)b^2_1(m,\chi)\,C_3(x)}\right]Y_{2-2}(\mathbf{\hat{x}};\mathbf{\hat{n}_{12}}),
	\end{align}
	where $C_1,\,C_2$ and $C_3$ are functions of $x$ obtained after performing the summation and integration in Eq.(\ref{final_result_1}) and $A$ is a pure numerical constant (see Appendix \ref{App:Derivation_details} for details). We use the linear bias factor given by \cite{2010ApJ_Tinker_bias} and for the second-order bias we use the analytic fitting formula provided by  \cite{2016JCAP_FabianBias}. 
	From Eq.(\ref{final_result_1}), we notice that the ensemble-averaged value of the polarisation depends on the spatial separation between the clusters. The dependence on cosmology enters via the Hubble parameter, growth factor, and the growth rate as well as through the linear matter power spectrum $P(k)$. The signal also depends on the first and second-order bias parameters. We also note that the ensemble averaging has been done for a fixed effective mass of a cluster pair and a fixed angle between the line of sight direction $\mathbf{\hat{n}_{12}}$ and cluster separation vector $\mathbf{x}$. 
\subsection{Effective optical depth \label{effective-opt_depth}}
To estimate the pairwise signal, we first need to measure the mean optical depth for a cluster pair. The optical depth depends on the electron density profile of the ICM, which we model using a spherically symmetric $\beta$ profile. The electron number density can be related to the spherical over density mass $M_{500c}$ as,
\begin{align}
	\label{norm_ne}
	M_{500c}=\frac{\Omega_m}{\Omega_b}\,1.14\,m_p\int_{0}^{R_{500c}}n_e(r)r^2dr
\end{align}
where $m_p$ is the mass of the proton, $\Omega_m$ and $\Omega_b$ are respectively the ratio between matter and baryon density to the critical density at redshift 0. We use Eq.(\ref{norm_ne}) to normalise the electron number density. More details about the electron density profile are given in Appendix \ref{App:cluster_profile}. The optical depth along a line of sight through the cluster which is at a semi-vertical angle of $\Phi$ from the line joining the observation point and the cluster centre  is given by \cite{2017ApJ_Fender},
	\begin{align}
		\label{tau}
		\tau(\Phi)=2\sigma_{\mathrm{T}}\int^{R_\mathrm{max}}_{\Phi d_A(\chi)}d\alpha \,n_e(\alpha)\left(\frac{\alpha}{\sqrt{\alpha^2-\Phi^2 d^2_A(\chi)}}\right),%=2\sigma_{\mathrm{T}}\,\tau_{\mathrm{LOS}}(\Phi),
	\end{align}
where $d_A(\chi)$ is the physical angular diameter distance to the cluster and $R_\mathrm{max}$ is the maximum  physical radius till which we integrate.  The variable $\alpha$ denotes the radial distance from the cluster centre. Averaging over all the line of sight directions, we get the effective mean optical depth for a cluster pair. 
	\begin{align}
		\label{tau_effective}
		\tau_{\mathrm{eff}}=\frac{2\pi\int_{0}^{\Phi_\mathrm{max}}d\Phi\,\Phi\,\tau(\Phi)}{\pi\Phi^2_\mathrm{max}}.
	\end{align}
where $\Phi_\mathrm{max}=R_\mathrm{max}/d_A(\chi)$. The value of $R_\mathrm{max}$ depends of the size of the cluster and the beam size of the instrument used in measuring the pairwise signal.  The angular size of the cluster is given by, $\sigma_{\mathrm{cluster}}=\pi \Phi^2_\mathrm{max}$. If the angular size of a cluster is more than the angular size of the beam (beam area), then the cluster will be resolved. If the cluster is resolved then we can we choose $\Phi_\mathrm{max}$ or equivalently $R_\mathrm{max}$ to maximize the signal. We found $R_\mathrm{max}=3.5\,R_{500c}$ yields the best result. If the cluster is unresolved, $\Phi_\mathrm{max}$ is equal to the angular radius of the instrument beam. Once, we calculate the mean effective optical depth, it is possible to evaluate the formula given in Eq.(\ref{final_result_Mz_form}). 
\section{Forecast for the pairwise pkSZ effect\label{Obs_connect}}
In observations, we have only one realization of the Universe. In order to obtain the pairwise pkSZ signal we average over many cluster pairs which are at a fixed separation from each other. There are some important points to note while doing the averaging. The polarisation vector of different clusters will be oriented along different directions on the plane of the sky. Therefore, we need to set a convention for the coordinate system on the plane of the sky for all pairs. The convention should be such that the polarisation from different clusters adds coherently when they have the same transverse velocity direction relative to their pair separation vector projected on the plane of the sky. This is important as $\mathcal{Q}+i\mathcal{U}$ is not invariant under a coordinate rotation. Expanding the spherical harmonics $Y_{2-2}(\mathbf{\hat{x}};\mathbf{\hat{n}_{12}})$  in Eq.(\ref{final_result_1}) we get \cite{durrer2020cosmic},
\begin{align}
	\label{sph_harm}
	Y_{2-2}(\mathbf{\hat{x}};\mathbf{\hat{n}_{12}})=\frac{1}{4}\sqrt{\frac{15}{2\pi}} \sin^2 \theta \exp(-2i\phi),
\end{align}
where $\theta$ is the angle between the cluster pair separation vector and the line of sight direction given by $\cos^{-1}(\mathbf{\hat{n}_{12}}\cdot \mathbf{\hat{x}})$ and $\phi$ is the angle between the projection of separation vector on the plane of the sky and the x-axis of our coordinate system. We have the freedom to rotate the coordinate system for each cluster pair such that $\phi=0$. This sets the convention of our coordinate system. After rotation, so that the separation vector is along the x-axis, we add the $\mathcal{Q}$ and $\mathcal{U}$ parameters from the two clusters and find the average for many such cluster pairs. We see that with this convention $P_{\mathrm{pairwise}}$ becomes real, i.e. pure $\mathcal{Q}$ as expected. The $\mathcal{U}$ parameter provides a null test and an estimate of the error on $\mathcal{Q}$. Therefore, from Eq.(\ref{final_result_Mz_form}), we get,
\begin{align}
	\label{final_result_Mz_form2}
	P_{\mathrm{pairwise}}(x \,|m,\chi,\theta)=A'\left[(D^2Hfa)^2\right]\hspace{-0.15cm}(\chi)\,\tau_{\mathrm{eff}}(m,\chi)\, \frac{b_1(m,\chi)\,C_1(x)+b_2(m,\chi)\,C_2(x)}{1+D^2(\chi)\,b^2_1(m,\chi)\,C_3(x)}\sin^2\theta,
\end{align}
where $A'=\frac{1}{4}\sqrt{\frac{15}{2\pi}}\,A$. The polarisation signal spectrum is a linear combination of the y-type spectrum and the differential blackbody spectrum given by, $2g(x)+\frac{1}{2}y(x)$, where $g(x)=\frac{xe^x}{e^x-1}$ is the differential blackbody spectrum, $y(x)=\frac{xe^x}{(e^x-1)}\left(x\frac{e^x+1}{e^x-1}-4\right)$ is the y-type distortion spectrum, and $x=\left({h\nu}/{k_BT}\right)$ is the dimensionless frequency, $\nu$ is the frequency of the polarised radiation, $h$ is the Planck constant, $k_B$ is the Boltzmann constant, and $T$ is the average monopole temperature of the CMB. Note that $g(x)$ is also the spectrum of the primordial CMB anisotropies for all the CMB experiments which make a differential measurement of the CMB. We can observe both $g(x)$ and $y(x)$ parts of the spectrum, but only the y-type distortion can be distinguished from the other primary and secondary sources of CMB distortion, provided we can measure the polarisation signal with multiple frequency channels. Thus, in order to compare with different CMB polarisation signals, we will present the pairwise signal in temperature units using a relation valid for the Rayleigh-Jeans part of the spectrum, $\Delta T=$2(y-amplitude)$T_{\mathrm{CMB}}$ for the y-distortion, where $T_{\mathrm{CMB}}=T(z=0)$ is the average CMB temperature today.\\\\
	\begin{figure}%[H]
	\centering
	\begin{subfigure}{0.84\textwidth}
		%	\vfill
		\centering
		\includegraphics[width=0.84\linewidth]{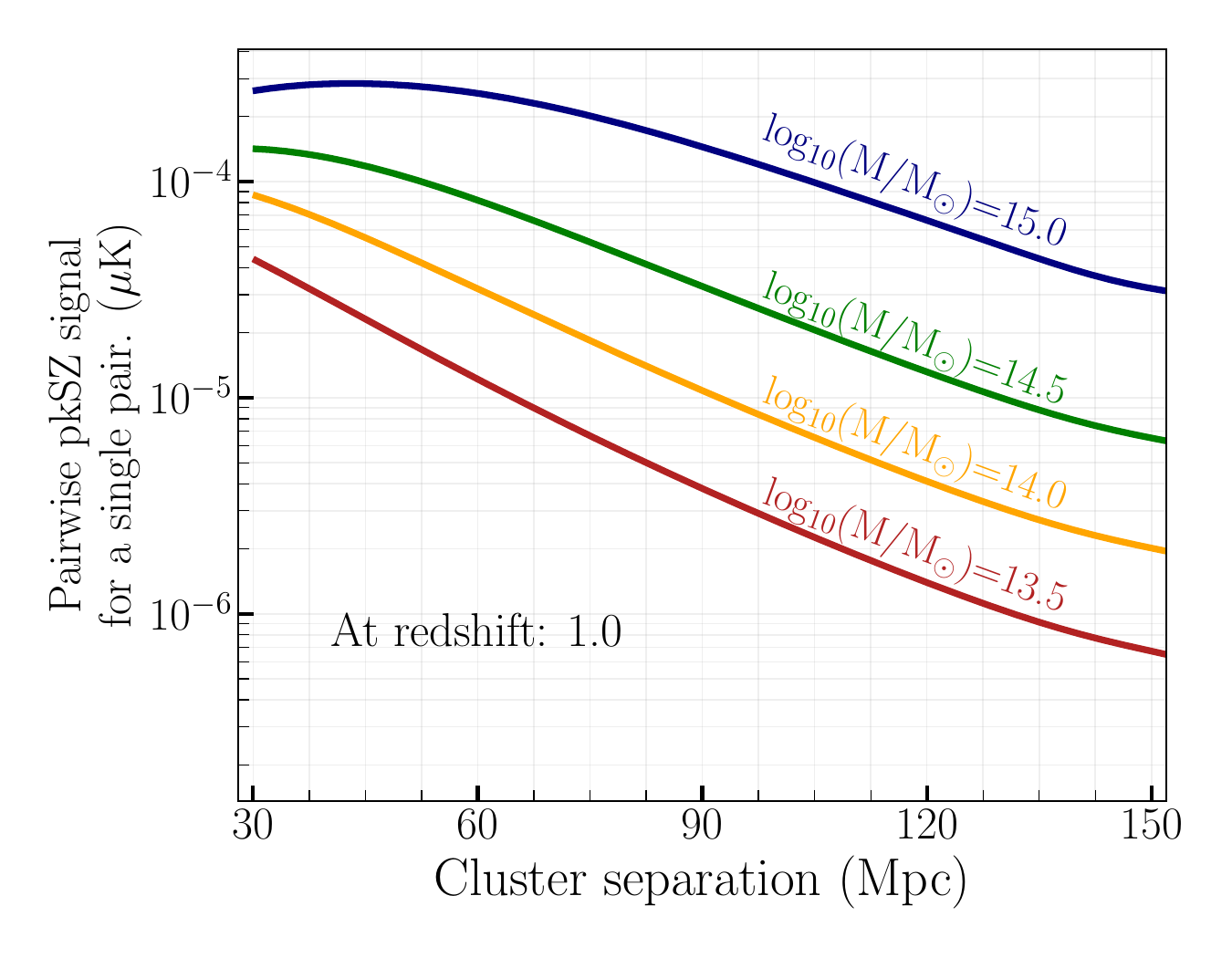}
	\end{subfigure}
	\caption{The y-axis shows the amplitude of the pairwise pkSZ effect (in $\mu$K). The 3-D cluster separation (in comoving Mpc units) is given on the x-axis. The coloured lines represent different masses. As expected, low mass cluster has smaller optical depth and therefore smaller polarisation signal.}
	\label{fig:signal and s_n}
\end{figure}\begin{table}%[H]
\centering
	\begin{tabular}{|c|c|}
		\hline
		$\log_{10}\left(M^{500c}_{\mathrm{min}}/M_\odot\right)$ & \begin{tabular}[c]{@{}c@{}}$\mathrm{N_{cl}}(M_{\mathrm{min}}),\;z_\mathrm{max}=3$\end{tabular} \\ \hline
		13.5                                & 4174260                                                            \\ \hline
		14                                & 396444                                                             \\ \hline
		14.5                               &18575                                                                   \\ \hline
		15                              & 229                                        \\ \hline
	\end{tabular}
	\caption{Average number of clusters available in the observable Universe above a certain mass. We used Eq.(\ref{cluster_number}) and mass function given by \cite{2008_Tinker} to obtain these numbers.}
	\label{tab:cl_no}
\end{table}The formula in Eq.(\ref{final_result_Mz_form2}) is for a particular average mass, comoving distance, and a given angle $\theta$ between the cluster pair and the line of sight direction. In observations, however, different cluster pairs will have different values for all these variables. Averaging $\sin^2\theta$ in Eq.(\ref{final_result_Mz_form2}) over all possible values of $\theta$ between $0$ and $\pi$ yields a factor of $1/2$. In figure \ref{fig:signal and s_n}, we have plotted the signal, as given in Eq.(\ref{final_result_Mz_form2}), for a single cluster pair of a given average mass as a function of spatial separation between cluster pairs, for the average value of $\sin^2\theta$. Since a cluster pair with a smaller average mass has a smaller optical depth, the resultant polarisation signal is also smaller. Therefore, if we include more small mass halos in the average, the signal decreases. To stay in the regime of quasi-linear theory as well as to neglect the effect of the finite size of the cluster, the minimum separation that we consider is $\simeq10\,\mathrm{Mpc}$ in comoving units.  Furthermore, averaging cluster pairs will also lead to averaging over different masses and comoving distances. Therefore, we must average Eq.(\ref{final_result_Mz_form2}) over mass and comoving distance or redshift distribution of the cluster pairs. The average over different cluster pair average masses and redshift depend on the properties of the cluster catalog being used. Catalogs from different surveys will have different selection functions. We expect cluster catalogs with a few hundred thousand clusters will be available in the near future x-ray survey such as eROSITA \cite{2012_erosita} and ground-based CMB surveys such as CMB-S4  \cite{2016_cmbs4}. Let us first consider for simplicity the ideal case, where the catalog has all clusters above a certain mass threshold $M_{\mathrm{min}}$ up to $z_\mathrm{max}=3$. We use four $M_{\mathrm{min}}$ values which are given by $\log \left(M_{500c}/M_\odot\right)=13.5,14.0,14.5,$ and $15.0$. We use $M=M_{500c}$, the mass within radius $R_{500c}$ where the average overdensity is $ 500$ times the critical density at a given redshift. The number of available clusters in the Universe above a certain mass $\mathrm{N_{cl}}(M_{\mathrm{min}})$ and in a certain redshift range is given by integrating the mass function \cite{2008_Tinker},
\begin{align}
\label{cluster_number}
\mathrm{N_{cl}}(M_{\mathrm{min}})&=4\pi\int_{0}^{z_\mathrm{max}}dz\frac{\chi^2(z)}{H(z)}\int_{M_{\mathrm{min}}}^{\infty}dm\frac{dn}{dm}\nonumber\\\nonumber\\
&=\int_{0}^{z_\mathrm{max}}\int_{M_{\mathrm{min}}}^{\infty}\left[4\pi \frac{\chi^2(z)}{H(z)}\frac{dn}{dm}\right]dm\,dz.
\end{align}
We can define the joint distribution function, 
\begin{align}
	\label{cl_dist}
p(m,z)=4\pi \frac{\chi^2(z)}{H(z)}\frac{dn}{dm},
\end{align}
with which the clusters are distributed in the mass-redshift space. We use this distribution function to sample clusters for our mock catalogs. We tabulate $\mathrm{N_{cl}}(M_{\mathrm{min}})$ in table \ref{tab:cl_no}. To calculate the signal for different mass thresholds, we sample the average mass and redshift of cluster pairs from Eq.(\ref{cl_dist}). We also sample $\theta$ from a uniform distribution between $0$ and $\pi$. Therefore, from Eq.(\ref{final_result_Mz_form2}), we get,
 \begin{align}
 \hspace*{-0.1cm}	
	 	\label{signal1}
	 	P_{\mathrm{pairwise}}(x)&=A'\sum_{i}\,\frac{w_i\left[(D^2Hfa)^2\right]\hspace{-0.15cm}(\chi_i)\,\tau_{\mathrm{eff}}(m_i,\chi_i)\left[b_1(m_i,\chi_i)\,C_1(x)+b_2(m_i,\chi_i)\,C_2(x)\right]}{1+D^2(\chi_i)\,b^2_1(m_i,\chi_i)\,C_3(x)}\sin^2\theta_i\nonumber\\
	 	&=\sum_{i}w_i\,P_{\mathrm{pairwise}}(x \,|m_i,\chi_i,\theta_i),
	 \end{align}
	 where the sum is over all cluster pairs we get from our mock catalog. The total number of cluster pairs is given by the Peebles-Hauser (natural) estimator \cite{1974ApJ_pH_estimator,2013A&A_estimator},
	 \begin{align}
	 	\label{DD}
	 	DD(x)=\left(1+(\bar{b}_1)^2\,\xi(x)\right)RR(x),
	 \end{align}
where $\xi(x)$ is the two-point correlation function and $\bar{b}_1$ mass-averaged linear bias factor of clusters at a given redshift. $RR$ denotes the total number of cluster pairs at a given separation in a randomly distributed sample, and $DD$ denotes the total number of cluster pairs in the presence of clustering. $RR(x)$ can be calculated easily as given in Appendix \ref{App:N_pair}. Therefore, knowing $RR(x)$ and $\xi(x)$, we can use Eq.(\ref{DD}) to find out the total number of cluster pairs available at a particular separation.  We give the average number of cluster pairs above a given mass cutoff in table \ref{tab:cl_pair}. 
%Eq.(\ref{signal1}) gives the theoretical expectation for the simplest estimator, where  $w_i=1/\sum_{\mathrm{N_{pairs}}}$ i.e. 
%\begin{align}
%	\label{simple_estimator}
%		\hat{P}^{}_{\mathrm{pairwise}}=\frac{1}{\mathrm{N_{pairs}}}\;\sum_{i}(P_{i1+}+P_{i2+})\Big|_{\mathrm{separation\,along\,x-axis}}.
%\end{align}
\begin{table}[]
	\centering
	\begin{tabular}{|c|lllll|}
		\hline
		\multirow{2}{*}{$\log_{10}\left(M^{500c}_{\mathrm{min}}/M_\odot\right)$} & \multicolumn{5}{l|}{\hspace{3.5cm}Separation bins (Mpc)}                                                                                                                                       \\ \cline{2-6} 
		& \multicolumn{1}{l|}{10-50}            & \multicolumn{1}{l|}{50-90}            & \multicolumn{1}{l|}{90-130}           & \multicolumn{1}{l|}{130-170}          & 170-210          \\ \hline
		13.5                                                                     & \multicolumn{1}{l|}{$1.58\times10^7$} & \multicolumn{1}{l|}{$6.78\times10^7$} & \multicolumn{1}{l|}{$1.63\times10^8$} & \multicolumn{1}{l|}{$3.04\times10^8$} & $4.84\times10^8$ \\ \hline
		14.0                                                                     & \multicolumn{1}{l|}{$2.99\times10^5$} & \multicolumn{1}{l|}{$1.08\times10^6$} & \multicolumn{1}{l|}{$2.55\times10^6$} & \multicolumn{1}{l|}{$4.75\times10^6$} & $7.52\times10^6$ \\ \hline
		14.5                                                                     & \multicolumn{1}{l|}{$1.92\times10^3$} & \multicolumn{1}{l|}{$5.02\times10^3$} & \multicolumn{1}{l|}{$1.12\times10^4$} & \multicolumn{1}{l|}{$2.09\times10^4$} & $3.25\times10^4$ \\ \hline
		15.0                                                                     & \multicolumn{1}{l|}{$1$}              & \multicolumn{1}{l|}{$2$}              & \multicolumn{1}{l|}{$4$}              & \multicolumn{1}{l|}{$7$}              & $11$             \\ \hline
	\end{tabular}
\caption{Average number of cluster pairs available in the observable Universe above a given lower mass cutoff of the clusters.}
\label{tab:cl_pair}
\end{table}We can now use Eq.(\ref{signal1}) to study the detectability of the pairwise pkSZ effect in future CMB experiments.
\subsection{Signal to noise ratio \label{s_to_n}}
In general, the source of noise can be from the instrumentation or from other components which can not be differentiated from the signal we want to observe. However, in the case of the pkSZ effect, given we can measure the polarisation in multiple frequency channels, we can differentiate the signal from other primary and secondary CMB components as well as foregrounds. This is due to the fact that the pkSZ is the dominant signal that is both polarised and has a y-type distortion in the frequency spectrum. Therefore, we will assume that the main source of noise is just the instrument noise in order to study the feasibility of detecting the signal in this idealised case. The sensitivity $\kappa$ of CMB experiments is usually specified in units of $\mu \mathrm{K\,arcmin}$. If a cluster is resolved\footnote{ At a beam of $1$arcmin FWHM (Full Width at Half Maximum), almost all the clusters will be resolved. But, if a cluster is not resolved, we will have to replace the area of the cluster with the area of the beam when we are calculating the signal, i.e. $\sigma_{\mathrm{cluster}}=\pi\Phi^2_\mathrm{max}\equiv\sigma_{\mathrm{beam}}$ in Eq.(\ref{tau_effective}). In this case, we will lose signal as we have to average the optical depth over an area which includes regions outside the cluster, where the electron number density is very close to zero. Also, while calculating the noise variance from a cluster, we will need to use $\sigma_{\mathrm{cluster}}\equiv\sigma_{\mathrm{beam}}$ in Eq.(\ref{noise_cl}). \label{footnote_1}} 
and has an angular area $\sigma_{\mathrm{cluster}}$ in $\mathrm{arcmin}^2$, the noise variance when averaged over the cluster is given by,
\begin{align}
	\label{noise_cl}
\left(	\mathcal{N}_\mathrm{cl}\right)^2= \frac{\kappa^2}{\sigma_{\mathrm{cluster}}}
\end{align}
Since we simply add the $\mathcal{Q}$ and $\mathcal{U}$ parameters from the two clusters in the pair, in the pairwise estimator, the noise variance for the cluster pair is $2\,\mathcal{N}^2_\mathrm{cl}$. When we do weighted average over $\mathrm{N_{pairs}}$ cluster pairs to measure the pairwise pkSZ signal, the resultant noise for our measurement becomes,
\begin{align}
	\label{noise}
	\mathcal{N}= \sqrt{2\,\sum_{i}\,w^2_i\left(\mathcal{N}^{\,i}_{\mathrm{cl}}\right)^2}
\end{align} 
where the sum is over all distinct cluster pairs. Using Eq.(\ref{signal1}) and Eq.(\ref{noise}) we can estimate the signal-to-noise ratio,
\begin{align}
	\label{s_n_main}
	\left(	\mathcal{\frac{S}{N}}\right)(x)=\frac{P_{\mathrm{pairwise}}(x)}{\mathcal{N}}.
\end{align}
\begin{figure}%[H]
	%\hspace*{-0.6cm}
	\centering
	\begin{subfigure}{0.84\textwidth}
		%	\vfill
		\centering
		\includegraphics[width=0.84\linewidth]{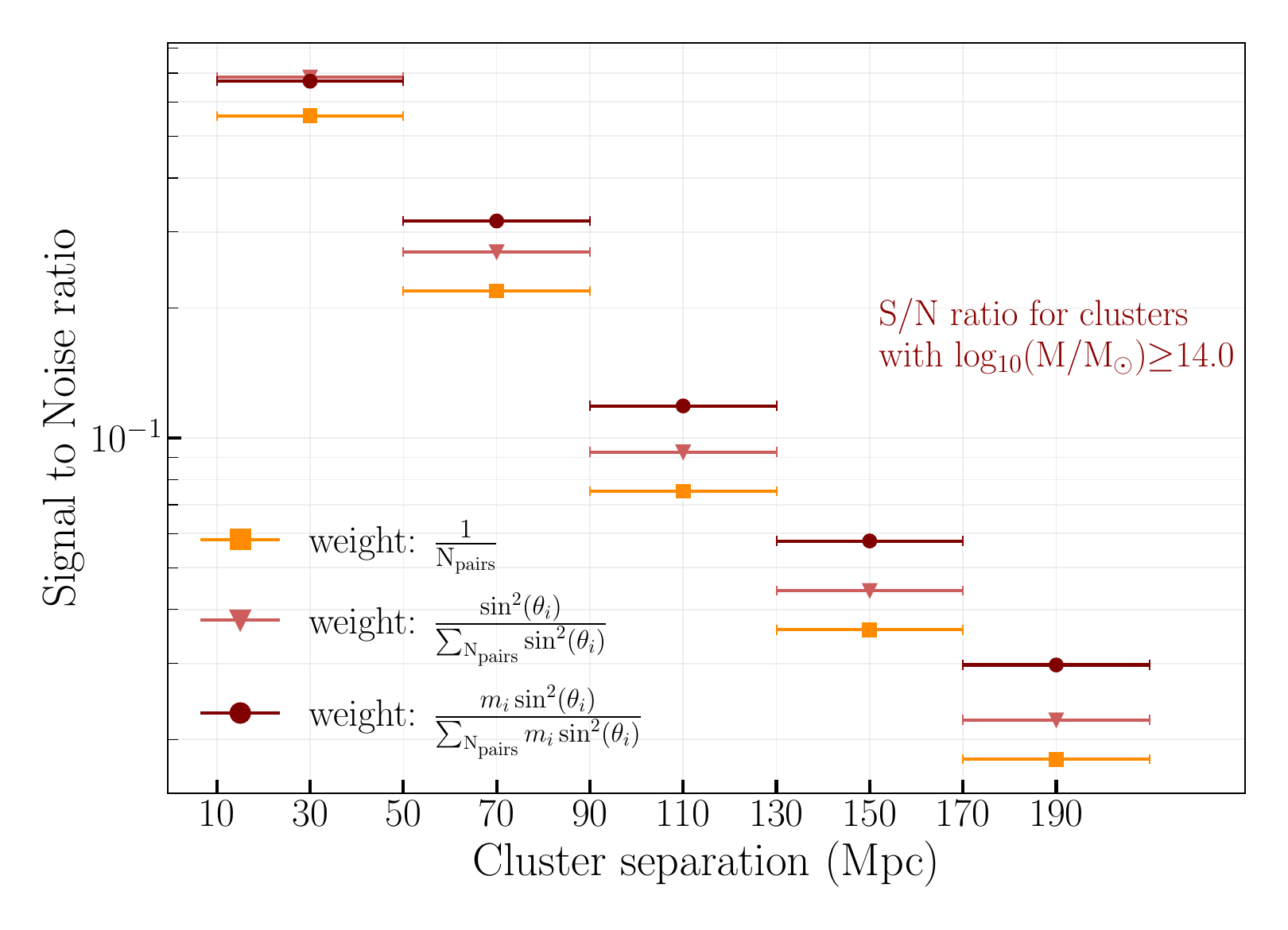}
	\end{subfigure}%
	\caption{Signal to noise ratio for different choice of weights in the pairwise estimator for an ideal survey where we detect all the clusters with $\log_{10}\left(M^{500c}_{\mathrm{min}}/M_\odot\right)\geq14$. The horizontal bars indicate the binning in the x-variable (separation distance).} 
	\label{fig:m_14_diff_w}
\end{figure}We use $\kappa=1\mu \mathrm{K\,arcmin}$ and a beam area $\sigma_{\mathrm{beam}}=1\mathrm{arcmin}$, similar to current planned or proposed CMB experiments, for our fiducial case. We can calculate the number of cluster pairs we require for a survey of given sensitivity to observe the pairwise pkSZ effect at a $\mathcal{S/N}=1$, using the scaling of $\mathcal{S/N}$ with sensitivity and $\mathrm{N_{pairs}}$, 
\begin{align}
	\label{calc_Npair}
	\mathrm{N_{pairs}}(\kappa)\Big|_{\mathcal{S/N}=1}=\mathrm{N_{pairs}}(1\mu\mathrm{Karcmin})\Big|_{\mathcal{S/N}=1}\left(\frac{\kappa}{1\mu\mathrm{Karcmin}}\right)^2.
\end{align}
Thus, an improvement in sensitivity by a factor of $10$, would result in a reduction in the number of cluster pairs needed by a factor of $100$.
\subsection{Comparison of different weight choices for the estimator\label{s_n_weighted}}
We want to choose the weights $w_i$ for each cluster pair so that the sign-to-noise ratio is maximized. The simplest choice of uniform weighting, $w_i=1/\mathrm{N_{pairs}}$, is sub-optimal. We would like to give higher weight to cluster pairs with higher signal and down-weight lower signal-to-noise ratio cluster pairs.  For cluster pairs with same mass and separation, the signal-to-noise ratio will be higher for pairs whose separation vectors are almost aligned with the plane of the sky, i.e. $\theta\simeq\pi/2$. Also, we want to give more weights to higher mass clusters, which have higher optical depth, resulting in a higher polarisation signal. Using optimization procedure as the linear kSZ effect \cite{2022MNRAS_DESIgalaxyclustersandPlanck}, since our signal is $\propto\sin^2\theta$, we get the $\hat{P}^{(1)}_{\mathrm{pairwise}}$ estimator,
\begin{align}
\label{weighted_estimator_1}
	&\hat{P}^{(1)}_{\mathrm{pairwise}}(x)=\frac{1}{\sum_{i}\sin^2\theta_i}\;\sum_{i}\sin^2\theta_i(P_{i1+}+P_{i2+})\Big|_{\mathrm{separation\,along\,x-axis}}.
\end{align}
Including mass weighting also yields the following estimator,
\begin{align}
\label{weighted_estimator_2}
	\hat{P}^{(2)}_{\mathrm{pairwise}}(x)=\frac{1}{\sum_{i}m_i\sin^2\theta_i}\;\sum_{i}m_i\sin^2\theta_i(P_{i1+}+P_{i2+})\Big|_{\mathrm{separation\,along\,x-axis}}.
\end{align}
 \begin{figure}%[H]
	\centering
	\begin{subfigure}{0.84\textwidth}
		%	\vfill
		\centering
		\includegraphics[width=0.84\linewidth]{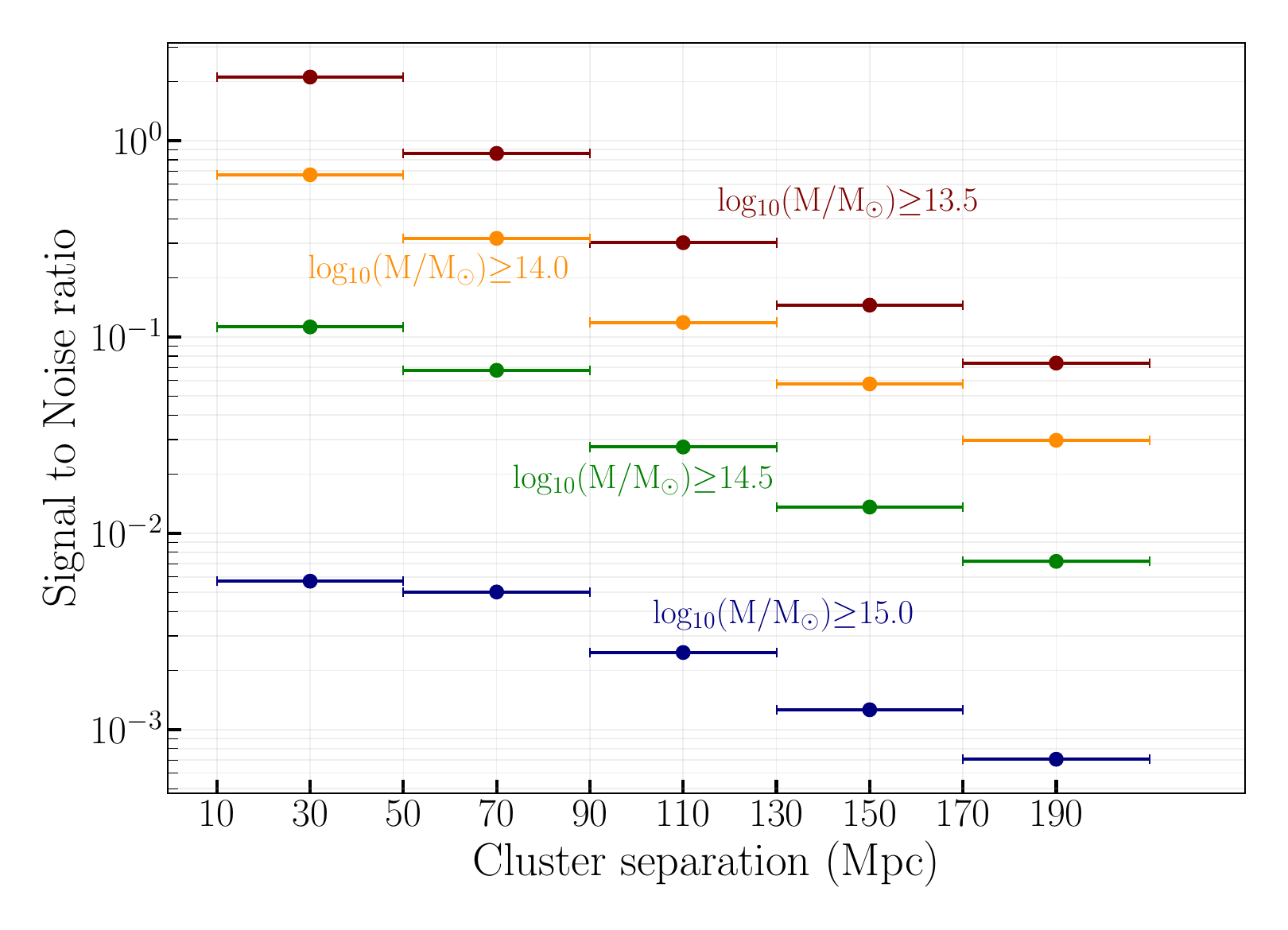}
	\end{subfigure}
	\caption{Signal to noise ratio for the optimal ($m\sin^2\theta$ weighted) pairwise estimator for ideal surveys where we detect all the clusters above a given lower mass cutoff.}
	\label{fig:diff_mass_2_weight}
\end{figure}We can obtain the theoretical expectation of these estimators by doing a weighted average of Eq.(\ref{final_result_Mz_form2}).
\begin{align}
\hspace*{-0.9cm}	
\label{signal2}
		P^{(1)}_{\mathrm{pairwise}}(x)&=\frac{1}{\sum_{i}\sin^2\theta_i}\sum_{i}\,P_{\mathrm{pairwise}}(x \,|m_i,\chi_i,\theta_i)\sin^2\theta_i \hspace{1cm}\mathrm{and}\\\nonumber
	P^{(2)}_{\mathrm{pairwise}}(x)&=\frac{1}{\sum_{i}m_i\sin^2\theta_i}\sum_{i}\,m_i\,P_{\mathrm{pairwise}}(x \,|m_i,\chi_i,\theta_i)\sin^2\theta_i.
\end{align}
In figure \ref{fig:m_14_diff_w}, we plot the signal-to-noise ratio for the pairwise pkSZ signal for cluster pairs above $\log \left(M_{500c}/M_\odot\right)=14$, as a function of separation between clusters in a pair. We can see that, the estimator resulting in simple averaging of polarisation signal from different cluster pairs is sub-optimal. Using $w=\sin^2\theta$ improves $\mathcal{S/N}$ and we get the best result from $m\sin^2\theta$ weighting. In figure \ref{fig:diff_mass_2_weight}, we plot the $m\sin^2\theta$ weighted estimator for the four different mock catalogs having different lower mass cutoffs for an ideal survey, where we can detect all clusters above a certain mass threshold with sensitivity of $1\mu$Karcmin. The number of cluster pairs available in the observable universe are given in table \ref{tab:cl_pair}. Since $\mathcal{S/N}\propto1/\kappa$, we see that with sensitivity of future experiment approaching sub-$\mu$Karcmin, we would be able to detect the pairwise pkSZ signal. We have also tried $m^\gamma$ weighting for $\gamma = 0.5, 0.75,1.5,$ and $2$, and find that these are sub-optimal compared to the linear weighting ($\gamma=1$).
	%\caption{We plot in figure (a), the amplitude of the pairwise pkSZ signal for the CMB-S4 wide survey. We have used the selection function given for the CMB-S4 wide survey for sampling clusters. The CMB-S4 wide survey will roughly observe $10^5$ clusters and will have a sensitivity of 1$\mu$K\,arcmin. We give a forecast for the number of cluster pairs required to reach a signal-to-noise of 1 in figure (b).}
	\begin{figure}%[H]
	\centering
	\begin{subfigure}{0.84\textwidth}
		%	\vfill
		\centering
		\includegraphics[width=0.84\linewidth]{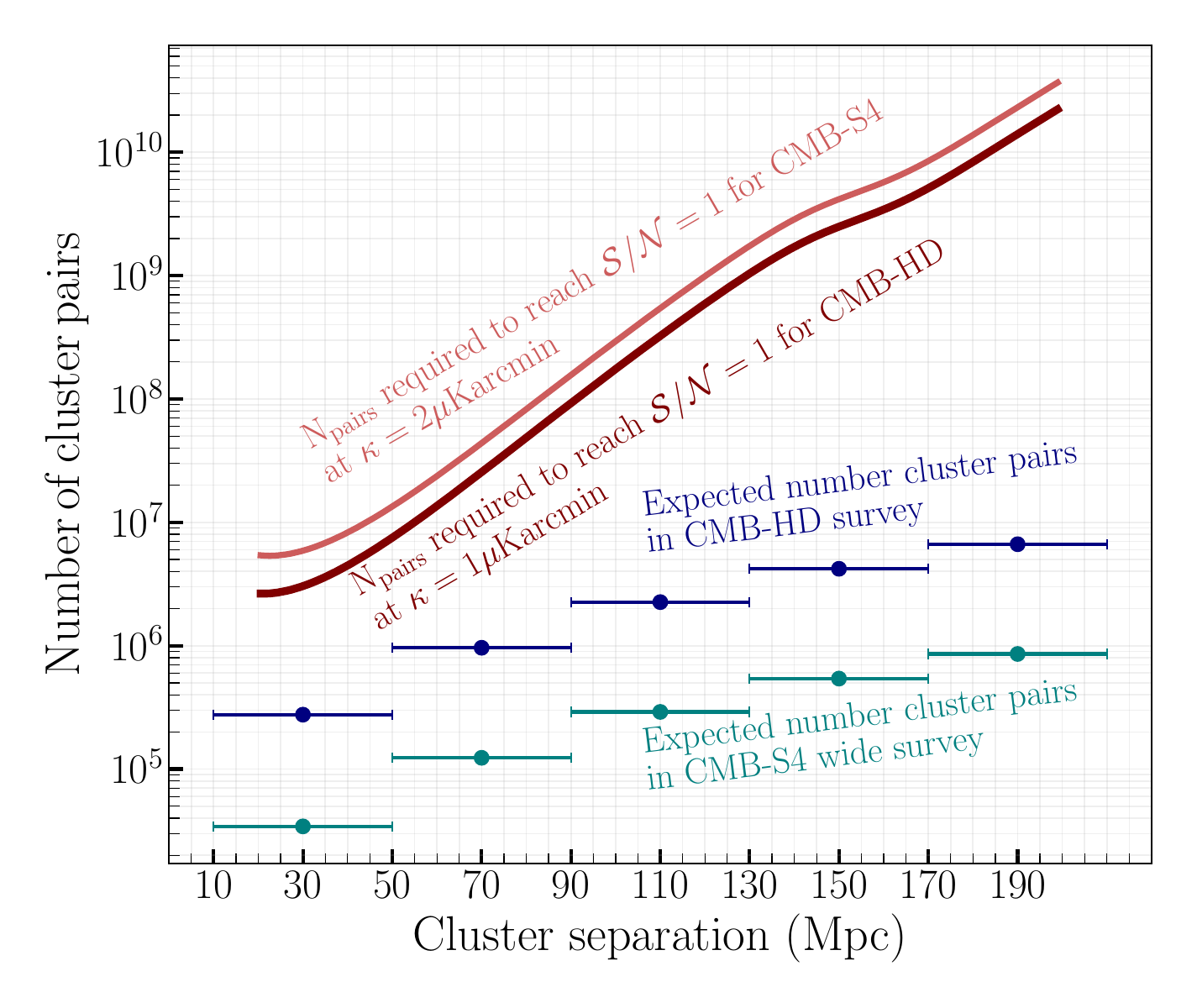}
	\end{subfigure}
	\caption{The ratio of the number of cluster pairs to the sensitivity required to observe the pairwise pkSZ effect at a signal-to-noise ratio of 1 for the optimal estimator ($w\propto m\sin^2\theta$) using CMB-S4 (in dashed red) and CMB-HD (in solid red) as a function of comoving 3-D separation cluster pair separation. We also give the number of cluster pairs expected in CMB-S4 (in teal) and CMB-HD (in blue).}
	\label{fig:cmb_s4_hd}
\end{figure}
\subsection{Forecast for CMB-S4 and CMB-HD surveys  \label{cmbs4}}
	The actual proposed CMB experiments, CMB-S4 and CMB-HD surveys, would cover $50\%$ of the sky and will have a redshift-dependent selection function. The CMB-S4 wide survey will detect approximately $10^5$ clusters and will have a sensitivity $\kappa\sim 2\mu \mathrm{K arcmin}$ for the polarised signals, while CMB-HD will detect about $4\times10^5$ clusters at a sensitivity of $1\mu \mathrm{K arcmin}$. We used the mass-redshift selection function given by \cite{2019_cmbs4_science_case,2022ApJ_raghunathan} for the CMB-S4 wide and CMB-HD to create a mock catalog of clusters with $10^5$ and $4\times10^5$ clusters respectively and calculate the pairwise signal for these catalogs. In figure \ref{fig:cmb_s4_hd}, we show our results in terms of the number of cluster pairs required to observe the pkSZ effect at $\mathcal{S/N}=1$ in these surveys. Note that this estimate assumes that the number of frequency channels will be sufficient to separate the pkSZ component with the y-type spectrum from other components, especially the polarised component with the blackbody spectrum. We also show the number of cluster pairs expected at each separation bin from the two surveys. Thus, we see that at the current sensitivity level, the pairwise pkSZ signal is just out of reach for the two surveys. A factor of 10 increase in the sensitivity or a cluster catalog with more clusters going to smaller $M_{\mathrm{min}}$ (see figure \ref{fig:diff_mass_2_weight}) will allow these surveys to observe the pairwise pkSZ effect.
		\begin{figure}%[H]
	\centering
	\begin{subfigure}{0.5\textwidth}
		%	\vfill
			%\vspace*{-0.cm}
			\hspace{-0.48cm}
		\centering
		\includegraphics[width=1.05\linewidth]{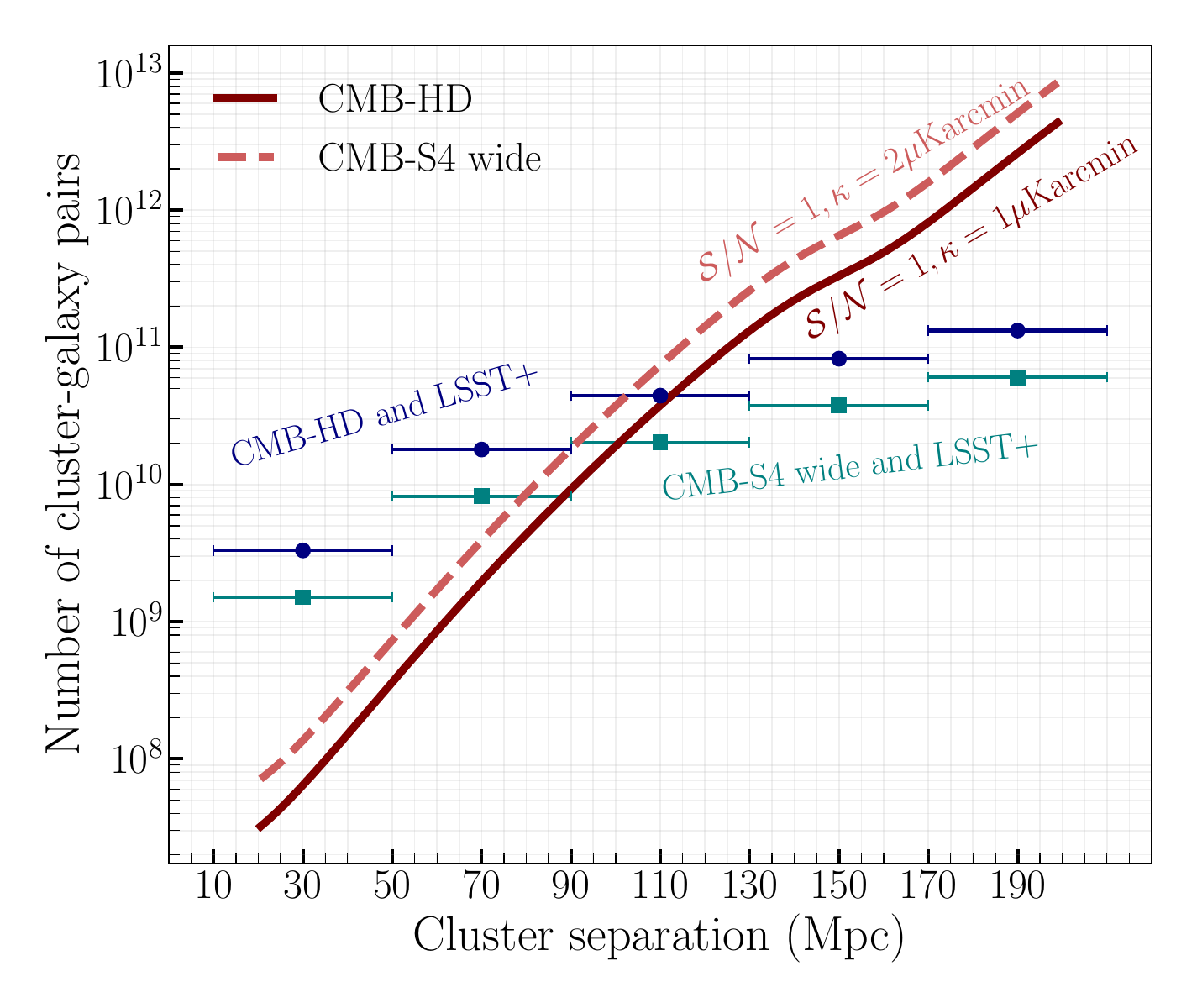}
		\caption{Number of cluster-galaxy pairs}
		\label{fig:lsst_ncl_1}
	\end{subfigure}%
	\begin{subfigure}{0.5\textwidth}
	%	\vfill
	\centering
	\includegraphics[width=1.05\linewidth]{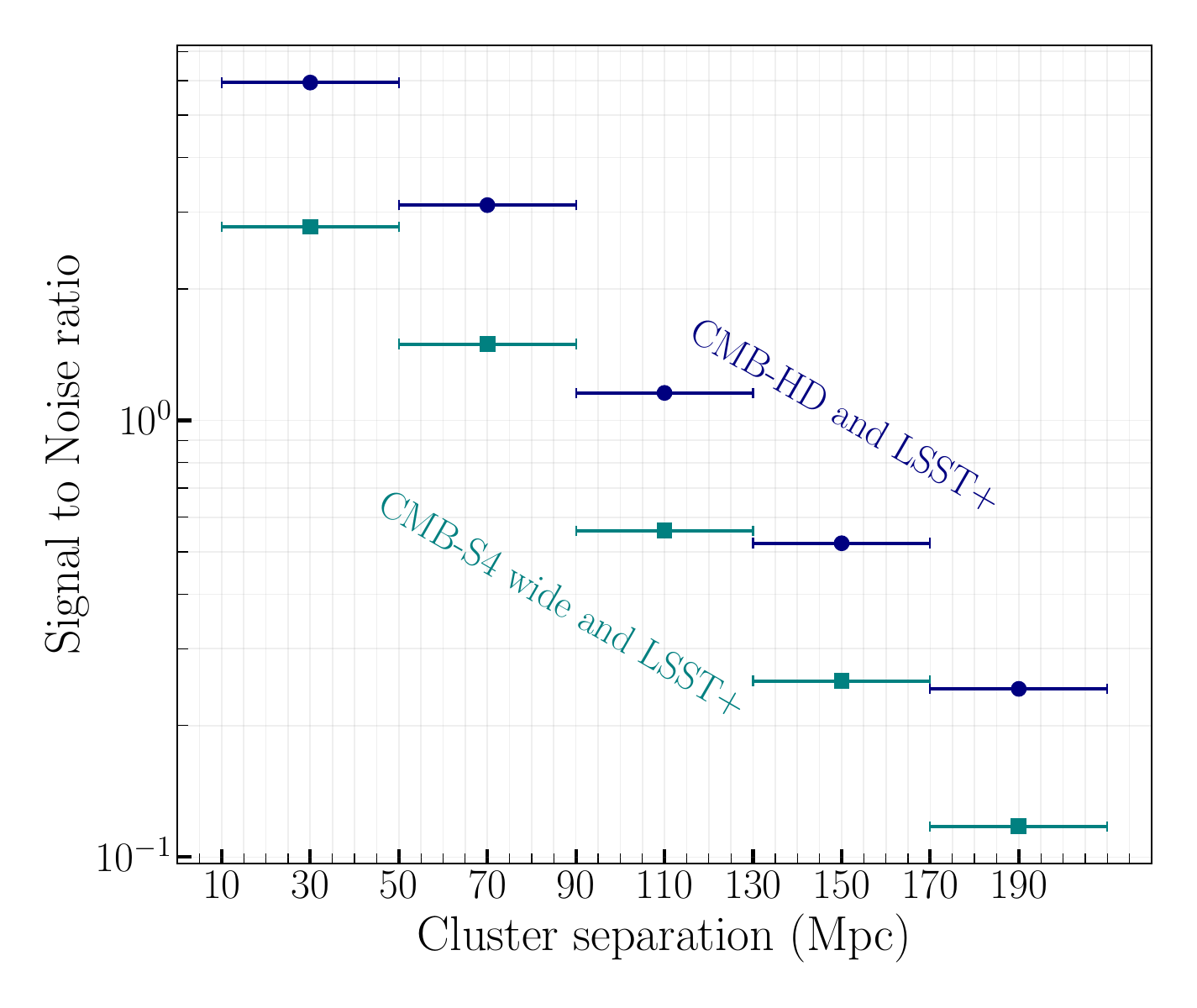}
	\caption{Signal-to-noise ratio}
	\label{fig:lsst_s_to_n}
\end{subfigure}
	\caption{Left panel: The number of cluster-galaxy pairs required to observe the cross-pairwise pkSZ effect at a signal-to-noise ratio of 1 using clusters from CMB-S4 wide and CMB-HD survey given in dashed and solid red lines respectively. In teal we provide the number of cluster-galaxy pairs we expect from CMB-S4 wide and LSST+ survey and in blue we have given the same for CMB-HD and LSST+ survey. The `+' symbol denotes that we will need spectroscopic redshift data of galaxies from future follow-ups of LSST cluster catalogs. Right panel: Signal-to-noise ratio for the cross-pairwise pkSZ effect with clusters from CMB-S4 wide and CMB-HD surveys with galaxies from LSST+ survey.}
	\label{fig:cmb_lsst}
\end{figure}
\section{Cross-correlating clusters with galaxies: Cross-pairwise pkSZ effect \label{cluster_galaxy}}
As discussed earlier, the chief objective of working with cluster pairs is to have a tracer of the direction of motion of clusters, in order to add polarisation from clusters coherently while stacking. However, clusters being rare objects are inefficient and noisy tracers of the large-scale gravitational potential and peculiar velocity fields. We should be able to improve the signal-to-noise ratio by cross-pairing clusters with a better tracer, i.e. galaxies. Thus, instead of pairing up two clusters, we can replace one of the clusters with a galaxy. The formalism developed for the cluster pairs remains unchanged. The Thomson optical depth of galaxies is very small and therefore, the polarisation signal is negligible. Also, since the galaxies are a better tracer of the underlying matter density field, the bias factors associated with galaxies are smaller than that of clusters. However, they are much more numerous than clusters. While there are $\mathcal{O}(10^6)$ galaxy clusters in the observable Universe, surveys like LSST \cite{2009_LSST} will detect $\mathcal{O}(10^9)$ galaxies. Future surveys can detect even more galaxies. We calculate the cross-pairwise pkSZ effect by pairing clusters from CMB-S4 wide or CMB-HD with galaxies from the LSST survey's ``Gold Sample", which will have a catalog of 4 billion galaxies. {\color{black}We provide a detail description about the estimation of the number of cluster galaxy pairs in Appendix \ref{App:N_pair}}. We note that the LSST survey will have a full sky overlap with the CMB-S4 survey. We need a precise 3-D position vector of the galaxies, at least better than a few Mpc, not just to know which bin the pair should fall in but also to calculate the $\sin^2\theta$ weight for stacking. Although we will have only photometric redshifts for galaxies from the LSST survey, future spectroscopic follow-ups are being planned \cite{2019_LSST_Spec}. We assume in our analysis that accurate 3-D positions from spectroscopic follow-ups will be available. {\color{black}It may be potentially possible to detect the pkSZ effect with just photometric redshifts as done in the case of pairwise kSZ \cite{2013_fabian_photo_z}}. We leave a detailed study of the impact of photometric redshift errors on the pkSZ signal for future work. For the cross-pairwise pkSZ effect, each pair has one cluster paired with one galaxy and we stack the CMB polarisation maps only at the position of the clusters. The only role of the galaxies is to determine the separation vector and hence the rotation angle by which the clusters should be rotated before stacking. We can modify Eq.(\ref{weighted_estimator_2}) to write the estimator for cluster-galaxy cross pairwise pkSZ effect, 
\begin{align}
		\hat{P}_{\mathrm{cl-gal}}(x)&=\frac{1}{\sum_{i}m_i\sin^2\theta_i}\;\sum_{i}m_i\sin^2\theta_i\,P_{i+}^{\mathrm{cl}}\Big|_{\mathrm{separation\,along\,x-axis}}.
\end{align}
We still require to orient the polarisation from clusters such that the separation vector between the cluster and a galaxy is along the x-axis of our coordinate system. Since, we are just stacking clusters, both the signal and noise variance get reduced by a factor of $1/2$ compared to the case where we paired clusters with clusters. Thus, we can get the corresponding expected value of the estimator by multiplying Eq.(\ref{signal1}) and by a factor of $1/2$,
 \begin{align}
	\hspace*{-0.6cm}	
	\label{signal_cl_gal}
	P_{\mathrm{cl-gal}}(x)&=\frac{A'}{2}\sum_{i}\,\frac{w_i\left[(D^2Hfa)^2\right]\hspace{-0.15cm}(\chi_i)\,\tau^{\mathrm{cl}}_{\mathrm{eff}}(m_i,\chi_i)\left[b^{\mathrm{gal}}_1\,C_1(x)+b^{\mathrm{gal}}_2\,C_2(x)\right]}{1+D^2(\chi_i)\,b^{\mathrm{gal}}_1\,b_1(m_i,\chi_i)\,C_3(x)}\sin^2\theta_i.
\end{align}
where $w_i=m_i\sin^2\theta_i/(\sum_{i}m_i\sin^2\theta_i)$. We have taken $b^{\mathrm{gal}}_1$ and $b^{\mathrm{gal}}_2$ to be the average galaxy bias factors for magnitude $m_r<24.5$ as given {\color{black}in Table 1 in ref. \cite{2022_gal_bias}}, $b^{\mathrm{gal}}_1=1.39$ and $b^{\mathrm{gal}}_2=0.034$. Similarly, we can get the noise by modifying Eq.(\ref{noise}) as,
\begin{align}
	\mathcal{N}= \sqrt{\,\sum_{i}\,w^2_i\left(\mathcal{N}^{\,i}_{\mathrm{cl}}\right)^2}.
\end{align} 
In figure \ref{fig:lsst_ncl_1}, we show the number of cluster-galaxy pairs required to observe the signal at $\mathcal{S/N}=1$ with LSST ``Gold Sample'' with spectroscopic data which we call LSST+ and using clusters from CMB-S4 and CMB-HD mocks in red dashed and solid lines respectively. In teal, we show the number of cluster-galaxy pairs we expect to get in each separation bin, when we select the clusters from the CMB-S4 wide survey and galaxies from LSST  ``Gold Sample". In navy blue, we show the same for clusters selected from the CMB-HD survey. We see that the number of expected cluster-galaxy pairs exceeds the number of required pairs in the separation range up to $\sim100$ Mpc. In figure \ref{fig:lsst_s_to_n}, we show the signal to noise ratio corresponding to CMB-S4 with LSST+ and CMB-HD with LSST+ surveys. It is evident from figure \ref{fig:lsst_ncl_1} and figure \ref{fig:lsst_s_to_n},that the pairwise pkSZ effect can be observed with high significance when cross-correlating CMB data with overlapping spectroscopic galaxy survey. We thus motivate a new science case for a spectroscopic follow-up of LSST galaxies. 
	\section{Conclusion \label{conclusion}}
	We present a new way to detect the polarised kinetic Sunyaev Zeldovich effect by pairing galaxy clusters with a tracer of the gravitational potential around the galaxy clusters. The theoretical formalism and the estimator are similar to the pairwise linear kSZ effect (a brief derivation is given in Appendix \ref{App:Review_kSZ}), albeit with some important differences. Since we are dealing with a 2-D transverse velocity field, we need to rotate the $\mathcal{Q}$ and $\mathcal{U}$ CMB maps at the location of clusters so that the signal adds coherently when stacking many clusters. The rotated frame with respect to which the $\mathcal{Q}$ and $\mathcal{U}$ parameters should be measured is defined by the separation vector between the clusters and a tracer of the matter field. We consider clusters of galaxies as well as galaxies themselves as the second element of the pair giving pairwise pkSZ effect and cross-pairwise pkSZ effect respectively. We show that averaging over many pairs results in a non-zero polarisation signal, provided a rotation is done so that the separation vector of each pair is aligned before stacking the clusters. Thus, we get an estimator that can be applied to the component separated (y-type component) $\mathcal{Q}$ and $\mathcal{U}$ parameter maps from future multi-frequency high-resolution CMB experiments. If we choose the x-axis along the cluster separation vector by rotating the patch of the sky containing the pair, for each pair, the stacked signal is a pure $\mathcal{Q}$. In this case, the $\mathcal{U}$ parameter provides a null test and an estimate of the error bar due to noise plus contamination accurately.\\\\
	We derive an analytical expression for the pairwise polarised kSZ effect from the first principles. The signal depends on the separation between the two members of a pair as well as on cosmological parameters like the growth factor, growth rate, the Hubble parameter and the linear matter power spectrum. The signal also depends on the linear and second-order halo bias parameters, being a higher-order effect compared to the unpolarised pairwise kSZ effect. We show that the proper choice of weights in the estimator increases the signal-to-noise ratio and present an optimised estimator to observe the pkSZ effect. For the catalogs in the near future from eROSITA as well as from the CMB-S4 wide survey, a factor of 10 increase in sensitivity compared to CMB-S4 and CMB-HD experiments, will enable us to observe the pairwise pkSZ signal from cluster pairs. We also show that cross-pairing clusters with galaxies from a spectroscopic survey like planned LSST follow-ups will enable detection with high significance even with the planned sensitivity of CMB-S4. We note that we should get a similar boost in $\mathcal{S/N}$ by cross-pairing clusters with galaxies in the linear kSZ effect also. The cross-pairwise linear kSZ effect using spectroscopic galaxy survey overlapping with the CMB experiments should be able to extract more information from data for the same reasons that result in a boost in $\mathcal{S/N}$ for the pkSZ effect. We leave a detailed study of the cross-pairwise linear kSZ effect for a separate paper. Another interesting possibility, which we have not explored in this paper, would be to pair clusters with voxels from future 21-cm or other line intensity mapping surveys. \\\\
	We have used perturbation theory for all analyses in this paper. In reality, there will be non-linear effects at smaller separations between members of the pairs (galaxies or clusters). We need to create mocks having 3-D positions of clusters and galaxies from N-body simulations and simulate CMB maps to make more accurate and quantitative predictions for specific cluster catalogs, CMB experiments, and galaxy surveys. Our present work shows that the pairwise and cross-pairwise pkSZ effect can be an important independent probe of cosmology and motivates more detailed studies.
	\acknowledgments
	This work is supported by the Department of Atomic Energy, Government of India, under Project Identification Number RTI 4002. We acknowledge the use of computational facilities of the Department of Theoretical Physics at Tata Institute of Fundamental Research, Mumbai. We want to specially thank Somnath Bharadwaj for his valuable comments about the connection between the BBGKY hierarchy and the ensemble averages. AKG is also thankful to Aseem Paranjape, Subhabrata Majumdar, Shadab Alam, Gilbert Holder, Ravi Sheth, Joseph Mohr, H\'ector Gil-Mar\'in and Anoma Ganguly for useful discussions.
	\appendix
	%\newpage
	\section{Derivation of the pairwise pkSZ effect\label{App:Derivation_details}}
	We start with the expression given in Eq.(\ref{contri_terms}).
	\begin{align}
		&\Big\langle (v_{1i}v_{1j}+v_{2i}v_{2j})(1+\delta^h_1)(1+\delta^h_2)\Big\rangle = b_1\Big\langle v_{1j}^{(1)}v_{1j}^{(1)}\delta_2^{(2)}\Big\rangle\:+b_1 \Big \langle v_{2i}^{(1)}v_{2j}^{(1)}\delta_1^{(2)}\Big\rangle+\nonumber\\
		&\hspace{3.5cm} \frac{b_2}{2} \Big \langle v_{1i}^{(1)}v_{1j}^{(1)}\left(\delta_2^{(1)}\right)^2\Big\rangle+ \frac{b_2}{2} \Big \langle v_{2i}^{(1)}v_{2j}^{(1)}\left(\delta_1^{(1)}\right)^2\Big\rangle\; + \mathrm{higher\, order\,terms}.
	\end{align}
As mentioned before, the first and the second term yields the same results, while the third and the fourth term gives the same result if we make the transformation, $\mathbf{k_1}\rightarrow-\mathbf{k_1}$ and $\mathbf{k_2}\rightarrow-\mathbf{k_2}$ in the Fourier space. Therefore, we can just evaluate the first and the third term. Let us begin at the first term. Using Eq.(\ref{delta1}), Eq.(\ref{delta2}), and Eq.(\ref{vel1}) we get
	\begin{align}
		\label{first_term}
		&\Big \langle v_{1i}^{(1)}v_{1j}^{(1)}\delta_2^{(2)}\Big\rangle=\int \frac{d^3\mathbf{k_1}d^3\mathbf{k_2}d^3\mathbf{k}}{(2\pi)^9}\int\frac{d^3\mathbf{q_1}d^3\mathbf{q_2}}{(2\pi)^3}\:\delta^3\left(\mathbf{q_1}+\mathbf{q_2}-\mathbf{k}\right)\exp\left(i (\mathbf{k_1}+\mathbf{k_2})\cdot\mathbf{x_1}\right)\nonumber\\
		&\hspace{1cm}D^4(Hfa)^2\exp\left(i \mathbf{k}\cdot\mathbf{x_2}\right)\frac{(ik_{1i})\,(ik_{2j})}{k_1^2\,k_2^2}\left[\frac{5}{7}+\frac{2}{7}\frac{(\mathbf{q_1}\cdot\mathbf{q_2})^2}{q_1^2\,q_2^2}+\frac{(\mathbf{q_1}\cdot\mathbf{q_2})}{2}\left(\frac{1}{q_1^2}+\frac{1}{q_2^2}\right)\right]\nonumber\\
		&\hspace{1cm}\Big\langle\delta_1^{(1)}(\mathbf{k_1})\delta_1^{(1)}(\mathbf{k_2})\delta_2^{(1)}(\mathbf{q_1})\delta_2^{(1)}(\mathbf{q_2})\Big\rangle.
	\end{align}
	Since, the perturbed variables are all Gaussian, we can use Isserlis theorem to evaluate the ensemble average. The non-zero terms contributing in the ensemble average is given by,
	\begin{align}
		\label{ensemble_first_term}
		\Big\langle\delta_1^{(1)}(\mathbf{k_1})\delta_1^{(1)}(\mathbf{k_2})\delta_2^{(1)}(\mathbf{q_1})\delta_2^{(1)}(\mathbf{q_2})\Big\rangle= \Big\langle\delta_1^{(1)}(\mathbf{k_1})\delta_2^{(1)}(\mathbf{q_1})\Big\rangle\Big\langle\delta_1^{(1)}(\mathbf{k_2})\delta_2^{(1)}(\mathbf{q_2})\Big\rangle+\nonumber\\
		\Big\langle\delta_1^{(1)}(\mathbf{k_1})\delta_2^{(1)}(\mathbf{q_2})\Big\rangle\Big\langle\delta_1^{(1)}(\mathbf{k_2})\delta_2^{(1)}(\mathbf{q_1})\Big\rangle.
	\end{align}
Since the second term gives identical contribution as the first term, we only need to evaluate the first term which can be written as,
	\begin{align}
		\Big\langle\delta_1^{(1)}(\mathbf{k_1})\delta_2^{(1)}(\mathbf{q_1})\Big\rangle\Big\langle\delta_1^{(1)}(\mathbf{k_2})\delta_2^{(1)}(\mathbf{q_2})\Big\rangle=(2\pi)^6\;P(k_1)P(k_2)\,\delta^3(\mathbf{k_1}+\mathbf{q_1})\,\delta^3(\mathbf{k_2}+\mathbf{q_2}),
	\end{align}
where $P(k)=P_{\mathcal{R}}(k)T^2(k)$ and $P_{\mathcal{R}}(k)$ is the power spectrum of the initial curvature perturbation. Therefore, from Eq.(\ref{first_term}) and Eq.(\ref{ensemble_first_term}) we get,
	\begin{align}
	b_1\Big\langle v_{1j}^{(1)}v_{1j}^{(1)}\delta_2^{(2)}\Big\rangle+ b_1\Big \langle v_{2i}^{(1)}v_{2j}^{(1)}\delta_1^{(2)}\Big\rangle=-&4\int \frac{d^3\mathbf{k_1}d^3\mathbf{k_2}}{(2\pi)^6}\exp\left(i (\mathbf{k_1}+\mathbf{k_2})\cdot\mathbf{x}\right)D^4(Hfa)^2\,b_1\nonumber\\
		&\hspace{-2.8cm}\frac{k_{1i}\,k_{2j}}{k_1^2\,k_2^2}\left[\frac{5}{7}+\frac{2}{7}\frac{(\mathbf{k_1}\cdot\mathbf{k_2})^2}{k_1^2\,k_2^2}+\frac{(\mathbf{k_1}\cdot\mathbf{k_2})}{2}\left(\frac{1}{k_1^2}+\frac{1}{k_2^2}\right)\right]P(k_1)P(k_2).
	\end{align}
Let us now look at the third term in Eq.(\ref{contri_terms}).
\begin{align}
	\label{third_term}
	& \Big \langle v_{1i}^{(1)}v_{1j}^{(1)}\left(\delta_2^{(1)}\right)^2\Big\rangle=D^4(Hfa)^2\int \frac{d^3\mathbf{k_1}d^3\mathbf{k_2}}{(2\pi)^6}\int\frac{d^3\mathbf{q_1}d^3\mathbf{q_2}}{(2\pi)^3}\exp\left(i (\mathbf{k_1}+\mathbf{k_2})\cdot\mathbf{x_1}\right)\nonumber\\
	 &\hspace{1cm}\exp\left(i (\mathbf{q_1}+\mathbf{q_2})\cdot\mathbf{x_2}\right)\frac{(ik_{1i})\,(ik_{2j})}{k_1^2\,k_2^2}\,\Big\langle\delta_1^{(1)}(\mathbf{k_1})\delta_1^{(1)}(\mathbf{k_2})\delta_2^{(1)}(\mathbf{q_1})\delta_2^{(1)}(\mathbf{q_2})\Big\rangle.
\end{align}
The ensemble average in this case is exactly similar to the expression in Eq.(\ref{ensemble_first_term}). Therefore, from  Eq.(\ref{third_term}) and Eq.(\ref{ensemble_first_term}) we get,
\begin{align}
	&\frac{b_2}{2} \Big \langle v_{1i}^{(1)}v_{1j}^{(1)}\left(\delta_2^{(1)}\right)^2\Big\rangle+ \frac{b_2}{2} \Big \langle v_{2i}^{(1)}v_{2j}^{(1)}\left(\delta_1^{(1)}\right)^2\Big\rangle=-4\,D^4(Hfa)^2\,\frac{b_2}{2}\, \frac{d^3\mathbf{k_1}d^3\mathbf{k_2}}{(2\pi)^6}\nonumber\\
	&\hspace{7cm}\exp\left(i (\mathbf{k_1}+\mathbf{k_2})\cdot\mathbf{x}\right) \frac{(k_{1i})\,(k_{2j})}{k_1^2\,k_2^2}\,P(k_1)P(k_2).
\end{align}
	From Eq.(\ref{main_b}) and Eq.(\ref{ensemble_defn_2}) we get,
	\begin{align}
		\label{eval_num_deno}
		\Big\langle v_{1i}v_{1j}+v_{2i}v_{2j}\Big\rangle_2\int \hat{n}'_i \hat{n}'_j Y_{2\lambda}^{*}(\mathbf{\hat{n}'})d\Omega_\mathbf{\hat{n}'}=\frac{\Big\langle (v_{1i}v_{1j}+v_{2i}v_{2j})(1+\delta^h_1)(1+\delta^h_2)\Big\rangle\int \hat{n}'_i \hat{n}'_j Y_{2\lambda}^{*}(\mathbf{\hat{n}'})d\Omega_\mathbf{\hat{n}'}}{\Big\langle (1+\delta^h_1)(1+\delta^h_2)\Big\rangle}.
	\end{align}
Let us first evaluate the numerator. We have already determined the relevant contribution in $\Big\langle (v_{1i}v_{1j}+v_{2i}v_{2j})(1+\delta^h_1)(1+\delta^h_2)\Big\rangle$. Therefore, we get,
	\begin{align}
		\label{expansion_point}
		& \Big\langle (v_{1i}v_{1j}+v_{2i}v_{2j})(1+\delta^h_1)(1+\delta^h_2)\Big\rangle\int \hat{n}'_i \hat{n}'_j Y_{2\lambda}^{*}(\mathbf{\hat{n}'})d\Omega_\mathbf{\hat{n}'}=-4D^4(Hfa)^2\,b_1\int d\Omega_\mathbf{\hat{n}'}Y_{2\lambda}^{*}(\mathbf{\hat{n}'})\nonumber\\
		&\int \frac{d^3\mathbf{k_1}d^3\mathbf{k_2}}{(2\pi)^6}\exp\left(i\mathbf{k_1}\cdot\mathbf{x}\right)\exp\left(i\mathbf{k_2}\cdot\mathbf{x}\right)\frac{(\mathbf{k_1}\cdot\mathbf{\hat{n}'})(\mathbf{k_2}\cdot\mathbf{\hat{n}'})}{k_1^2\,k_2^2}\Bigg[b_1\Bigg\{\frac{5}{7}+\frac{2}{7}\frac{(\mathbf{k_1}\cdot\mathbf{k_2})^2}{k_1^2\,k_2^2}+\nonumber\\
		&\hspace{4cm}\frac{(\mathbf{k_1}\cdot\mathbf{k_2})}{2}\left(\frac{1}{k_1^2}+\frac{1}{k_2^2}\right)\Bigg\}+\frac{1}{2}b_2\Bigg]P(k_1)P(k_2),
	\end{align}
	where $\mathbf{x}=(\mathbf{x_1}-\mathbf{x_2})$. We can rearrange Eq.(\ref{expansion_point}) in the following way,
	\begin{align}
		\label{expansion_point_2}
		&\Big\langle (v_{1i}v_{1j}+v_{2i}v_{2j})(1+\delta^h_1)(1+\delta^h_2)\Big\rangle\int \hat{n}'_i \hat{n}'_j Y_{2\lambda}^{*}(\mathbf{\hat{n}'})d\Omega_\mathbf{\hat{n}'}=-4D^4(Hfa)^2\int d\Omega_\mathbf{\hat{n}'}Y_{2\lambda}^{*}(\mathbf{\hat{n}'})\nonumber\\
		&\int \frac{d^3\mathbf{k_1}d^3\mathbf{k_2}}{(2\pi)^6}\exp\left(i(\mathbf{k_1}+\mathbf{k_2})\cdot\mathbf{x}\right)(\mathbf{\hat{k}_1}\cdot\mathbf{\hat{n}'})(\mathbf{\hat{k}_2}\cdot\mathbf{\hat{n}'})\sum^{2}_{q=0}G_{q}(k_1,k_2,b_1,b_2)\left(\mathbf{\hat{k}_1}\cdot\mathbf{\hat{k}_2}\right)^{q}P(k_1)P(k_2),
	\end{align}
	where,
	\begin{align}
		&G_{0}(k_1,k_2,b_1,b_2)=\left(\frac{5}{7}b_1+\frac{1}{2}b_2\right)\left(\frac{1}{k_1k_2}\right),\\
		&G_{1}(k_1,k_2,b_1,b_2)=\frac{1}{2}b_1\left(\frac{1}{k_1^2}+\frac{1}{k_2^2}\right),\hspace{1cm}\mathrm{and}\\
		&G_{2}(k_1,k_2,b_1,b_2)=\frac{2}{7}b_1\left(\frac{1}{k_1k_2}\right).
	\end{align}
	We expand the exponential and  and all the dot products in Eq.(\ref{expansion_point_2}) in terms of spherical harmonics and Bessel function \cite{varshalovich1988quantum}.
	
	\begin{align}
		&\mathbf{\hat{k}_1}\cdot\mathbf{\hat{n}'}	=\frac{4\pi}{3}\sum_{p_1}Y_{1p_1}^{*}(\mathbf{\hat{k}_1})Y_{1p_1}(\mathbf{\hat{n}'}),\\
		&\mathbf{\hat{k}_2}\cdot\mathbf{\hat{n}'}	=\frac{4\pi}{3}\sum_{p_2}Y_{1p_2}^{*}(\mathbf{\hat{k}_2})Y_{1p_2}(\mathbf{\hat{n}'}),  \;\;\mathrm{and}\\
		&\left(\mathbf{\hat{k}_1}\cdot\mathbf{\hat{k}_2}\right)^{q}=\sum_{l}\frac{q!}{(q-l)!!(q+l+1)!!}\sum_{p}Y_{lp}^{*}(\mathbf{\hat{k}_1})Y_{lp}(\mathbf{\hat{k}_2}).
	\end{align}
	In this formula the summation index $l$ assumes the values $l=0, 2, \cdots, q-2, q $ if q is even, and $l = 1, 3, \cdots,
	q-2, q$ if q is odd \cite{varshalovich1988quantum}.
	\begin{align}
		\exp(i\mathbf{k_1}\cdot\mathbf{x})=4\pi\sum_{L_1,M_1}i^{L_1}Y^{*}_{L_1M_1}(\mathbf{\hat{k}_1})Y_{L_1M_1}(\mathbf{\hat{x}})j_{L_1}(k_1x).\\
		\exp(i\mathbf{k_2}\cdot\mathbf{x})=4\pi\sum_{L_2,M_2}i^{L_2}Y^{*}_{L_2M_2}(\mathbf{\hat{k}_2})Y_{L_2M_2}(\mathbf{\hat{x}})j_{L_2}(k_2x).
	\end{align}
	\begin{align}
		\label{expansion_point_3}
		&\Big\langle (v_{1i}v_{1j}+v_{2i}v_{2j})(1+\delta^h_1)(1+\delta^h_2)\Big\rangle\int \hat{n}'_i \hat{n}'_j Y_{2\lambda}^{*}(\mathbf{\hat{n}'})d\Omega_\mathbf{\hat{n}'}=-4D^4(Hfa)^2 \sum_{p_1,p_2}\int d\Omega_\mathbf{\hat{n}'}Y_{2\lambda}^{*}(\mathbf{\hat{n}'})\nonumber\\
		&Y_{1p_1}(\mathbf{\hat{n}'})Y_{1p_2}(\mathbf{\hat{n}'})\int \frac{d^3\mathbf{k_1}d^3\mathbf{k_2}}{(2\pi)^6}\left(\frac{4\pi}{3}\right)^2\left(4\pi\right)^2\sum_{L_1,M_1}\sum_{L_2,M_2}\sum^{2}_{q=0}\sum_{l,p}i^{(L_1+L_2)}\frac{q!}{(q-l)!!(q+l+1)!!}\nonumber\\
		&Y_{L_1M_1}(\mathbf{\hat{x}})Y_{L_2M_2}(\mathbf{\hat{x}})G_{q}(k_1,k_2,b_1,b_2)Y_{1p_1}^{*}(\mathbf{\hat{k}_1})Y^{*}_{L_1M_1}(\mathbf{\hat{k}_1})Y_{lp}^{*}(\mathbf{\hat{k}_1})Y_{1p_2}^{*}(\mathbf{\hat{k}_2})Y^{*}_{L_2M_2}(\mathbf{\hat{k}_2})Y_{lp}(\mathbf{\hat{k}_2})\nonumber\\
		&j_{L_1}(k_1x)\,j_{L_2}(k_2x)P(k_1)P(k_2).
	\end{align}
	After expanding, we perform the angular integrals over the $\mathbf{\hat{n}'}$, $\mathbf{\hat{k}_1}$ and $\mathbf{\hat{k}_2}$ directions.
	\begin{align}
		\label{integral_n'}
	&	\int d\Omega_\mathbf{\hat{n}'}\left[Y_{2\lambda}^{*}(\mathbf{\hat{n}'})Y_{1p_1}(\mathbf{\hat{n}'})Y_{1p_2}(\mathbf{\hat{n}'})\right]=\sqrt{\frac{3\cdot3\cdot5}{4\pi}}\;(-1)^{\lambda}
		\left(\begin{array}{ccc}
			1& 1 & 2\\ 
			0& 0 & 0
		\end{array}\right)
		\left(\begin{array}{ccc}
			1& 1 & 2\\ 
			p_1& p_2 & -\lambda
		\end{array}\right),\\\nonumber\\
	&	\int d\Omega_{\mathbf{\hat{k}_1}}Y_{1p_1}^{*}(\mathbf{\hat{k}_1})Y^{*}_{L_1M_1}(\mathbf{\hat{k}_1})Y_{lp}^{*}(\mathbf{\hat{k}_1})=\sqrt{\frac{3(2L_1+1)(2l+1)}{4\pi}}
		\left(\begin{array}{ccc}
			1& L_1 & l\\ 
			0& 0 & 0
		\end{array}\right)
		\left(\begin{array}{ccc}
			1& L_1 & l\\ 
			p_1& M_1 & p
		\end{array}\right),\\\nonumber\\
	&	\int d\Omega_{\mathbf{\hat{k}_2}}Y_{1p_2}^{*}(\mathbf{\hat{k}_2})Y^{*}_{L_2M_2}(\mathbf{\hat{k}_2})Y_{lp}(\mathbf{\hat{k}_2})=\sqrt{\frac{3(2L_2+1)(2l+1)}{4\pi}}(-1)^p
		\left(\begin{array}{ccc}
			1& L_2& l\\ 
			0& 0 & 0
		\end{array}\right)
		\left(\begin{array}{ccc}
			1& L_2 & l\\ 
			p_2& M_2 & -p
		\end{array}\right).
	\end{align}
	Therefore, inserting the results of the above angular integrals in Eq.(\ref{expansion_point_3}), we get,
	\begin{align}
		\label{expansion_point_4}
		&\Big\langle (v_{1i}v_{1j}+v_{2i}v_{2j})(1+\delta^h_1)(1+\delta^h_2)\Big\rangle\int \hat{n}'_i \hat{n}'_j Y_{2\lambda}^{*}(\mathbf{\hat{n}'})d\Omega_\mathbf{\hat{n}'}=-4D^4(Hfa)^2 \sum_{p_1,p_2}\sqrt{\frac{3\cdot3\cdot5}{4\pi}}\;(-1)^{\lambda}
	\nonumber\\
		&\int \frac{dk_1dk_2k_1^2k_2^2}{(2\pi)^6}\left(\frac{4\pi}{3}\right)^2\left(4\pi\right)^2\sum_{L_1,M_1}\sum_{L_2,M_2}\sum^{2}_{q=0}\sum_{l,p}i^{(L_1+L_2)}\frac{q!}{(q-l)!!(q+l+1)!!}Y_{L_1M_1}(\mathbf{\hat{x}})Y_{L_2M_2}(\mathbf{\hat{x}})\nonumber\\
		&G_{q}(k_1,k_2,b_1,b_2)j_{L_1}(k_1x)\,j_{L_2}(k_2x)P(k_1)P(k_2)(-1)^p\sqrt{\frac{3(2L_1+1)(2l+1)}{4\pi}}\sqrt{\frac{3(2L_2+1)(2l+1)}{4\pi}}\nonumber\\
		&
		\left(\begin{array}{ccc}
			1& 1 & 2\\ 
			0& 0 & 0
		\end{array}\right)
		\left(\begin{array}{ccc}
			1& L_1 & l\\ 
			0& 0 & 0
		\end{array}\right)
		\left(\begin{array}{ccc}
			1& L_2& l\\ 
			0& 0 & 0
		\end{array}\right)
	\left(\begin{array}{ccc}
		1& L_1 & l\\ 
		p_1& M_1 & p
	\end{array}\right)
		\left(\begin{array}{ccc}
			1& L_2 & l\\ 
			p_2& M_2 & -p
		\end{array}\right)
	\left(\begin{array}{ccc}
		1& 1 & 2\\ 
		p_1& p_2 & -\lambda
	\end{array}\right).
	\end{align}
	We can further simplify by using the properties of summation of products of Wigner 3j symbols \cite{varshalovich1988quantum}. From Eq.(\ref{expansion_point_4}), we get,
	\begin{align}
		\label{3j_prop1}
		&\sum_{p_1,p_2,p}(-1)^{(\lambda+p)}
		\left(\begin{array}{ccc}
			1& 1 & 2\\ 
			p_1& p_2 & -\lambda
		\end{array}\right)
		\left(\begin{array}{ccc}
			1& L_1 & l\\ 
			p_1& M_1 & p
		\end{array}\right)
		\left(\begin{array}{ccc}
			1& L_2 & l\\ 
			p_2& M_2 & -p
		\end{array}\right)	\\
		=&
		\sum_{p_1,p_2,p}(-1)^{(p_1+p_2+p)}(-1)^{(l+L_1+1)}
		\left(\begin{array}{ccc}
			1& 2 & 1\\ 
			-p_1& \lambda & -p_2
		\end{array}\right)
		\left(\begin{array}{ccc}
			1& L_2 & l\\ 
			p_2& M_2 & -p
		\end{array}\right)	
		\left(\begin{array}{ccc}
			1& L_1 & l\\ 
			p& M_1 & p_1
		\end{array}\right)\\
		=&(-1)^{(L_1+1)}
		\left(\begin{array}{ccc}
			2& L_2 & L_1\\ 
			-\lambda& -M_2 & -M_1
		\end{array}\right)
		\left\{\begin{array}{ccc}
			2& L_2 & L_1\\ 
			l& 1& 1
		\end{array}\right\}.
	\end{align}
	Also, from the properties of spherical harmonics and 3j symbol we get \cite{varshalovich1988quantum}, 
	\begin{align}
		\label{3j_prop2}
		\sum_{M_1,M_2}
		\left(\begin{array}{ccc}
			2& L_2 & L_1\\ 
			-\lambda& -M_2 & -M_1
		\end{array}\right)
		Y_{L_1M_1}(\mathbf{\hat{x}})Y_{L_2M_2}(\mathbf{\hat{x}})=\sqrt{\frac{(2L_1+1)(2L_2+1)}{4\pi\cdot5}}
		\left(\begin{array}{ccc}
			L_1& L_2 & 2\\ 
			0& 0 & 0
		\end{array}\right)
		Y^{*}_{2\lambda}(\mathbf{\hat{x}}).
	\end{align}
	Therefore, putting the results of Eq.(\ref{3j_prop1}) and Eq.(\ref{3j_prop2}) in Eq.(\ref{expansion_point_4}), we finally get,
	\begin{align}
		\label{expansion_point_5}
		&\Big\langle (v_{1i}v_{1j}+v_{2i}v_{2j})(1+\delta^h_1)(1+\delta^h_2)\Big\rangle\int \hat{n}'_i \hat{n}'_j Y_{2\lambda}^{*}(\mathbf{\hat{n}'})d\Omega_\mathbf{\hat{n}'}=-4D^4(Hfa)^2\;Y^{*}_{2\lambda}(\mathbf{\hat{x}}) \sqrt{\frac{3\cdot3\cdot5}{4\pi}}\;
		\nonumber\\
		&\int \frac{dk_1dk_2k_1^2k_2^2}{(2\pi)^6}\left(\frac{4\pi}{3}\right)^2\left(4\pi\right)^2\sum_{L_1,L_2}\sum^{2}_{q=0}\sum_{l}i^{(L_1+L_2)}(-1)^{(L_1+1)}\frac{q!}{(q-l)!!(q+l+1)!!}\nonumber\\
		&G_{q}(k_1,k_2,b_1,b_2)j_{L_1}(k_1x)\,j_{L_2}(k_2x)P(k_1)P(k_2)\sqrt{\frac{3(2L_1+1)(2l+1)}{4\pi}}\sqrt{\frac{3(2L_2+1)(2l+1)}{4\pi}}\nonumber\\
		&
		\sqrt{\frac{(2L_1+1)(2L_2+1)}{4\pi\cdot5}}
		\left(\begin{array}{ccc}
			1& 1 & 2\\ 
			0& 0 & 0
		\end{array}\right)
		\left(\begin{array}{ccc}
			1& L_1 & l\\ 
			0& 0 & 0
		\end{array}\right)
		\left(\begin{array}{ccc}
			1& L_2& l\\ 
			0& 0 & 0
		\end{array}\right)
		\left(\begin{array}{ccc}
			L_1& L_2 & 2\\ 
			0& 0 & 0
		\end{array}\right)
		\left\{\begin{array}{ccc}
			2& L_2 & L_1\\ 
			l& 1& 1
		\end{array}\right\}.
	\end{align}
	In a similar way we can evaluate the denominator in Eq.(\ref{eval_num_deno}).
\begin{align}
	\Big\langle (1+\delta^h_1)(1+\delta^h_2)\Big\rangle=1+\langle\delta^h_1\delta^h_2\rangle=1+b_1^2\langle\delta_1^{(1)}\delta_2^{(1)}\rangle.
\end{align}
Using Eq.(\ref{delta1}) we get,
\begin{align}
	\label{deno_expansion}
	\Big\langle (1+\delta^h_1)(1+\delta^h_2)\Big\rangle&=1+D^2b^2_1\int\int\frac{d^3\mathbf{k_1}d^3\mathbf{k_2}}{(2\pi)^6}\exp\left(i\mathbf{k_1}\cdot\mathbf{x_1}\right)\exp\left(i\mathbf{k_2}\cdot\mathbf{x_2}\right)\Big\langle\delta_1^{(1)}(\mathbf{k_1})\delta_1^{(1)}(\mathbf{k_2})\Big\rangle\nonumber\\
	&=1+\frac{D^2b^2_1}{2\pi^2}\int dkk^2j_0(kx)P(k).
\end{align}
Using the results given in Eq.(\ref{expansion_point_5}) and Eq.(\ref{deno_expansion}) we can write the pairwise signal as,
	\begin{align}
		\label{p_pairwise1}
		&P_{\mathrm{pairwise}}(\mathbf{x},\mathbf{\hat{n}_{12}}|m,\chi)=-\frac{\sqrt{6}}{10}\left(-4D^4(Hfa)^2\right)
		\tau_{\mathrm{eff}}\sum_{\lambda}\,_2Y_{2\lambda}(\mathbf{\hat{n}_{12}})Y^{*}_{2\lambda}(\mathbf{\hat{x}}) \sqrt{\frac{3\cdot3\cdot5}{4\pi}}\;\nonumber\\
		&
	\int \frac{dk_1dk_2k_1^2k_2^2}{(2\pi)^6}\left(\frac{4\pi}{3}\right)^2\left(4\pi\right)^2\sum_{L_1,L_2}\sum^{2}_{q=0}\sum_{l}i^{(L_1+L_2)}(-1)^{(L_1+1)}\frac{q!}{(q-l)!!(q+l+1)!!}\nonumber\\
	&G_{q}(k_1,k_2,b_1,b_2)j_{L_1}(k_1x)\,j_{L_2}(k_2x)P(k_1)P(k_2)\sqrt{\frac{3(2L_1+1)(2l+1)}{4\pi}}
		\sqrt{\frac{3(2L_2+1)(2l+1)}{4\pi}}\nonumber\\
		&
		\sqrt{\frac{(2L_1+1)(2L_2+1)}{4\pi\cdot5}}
			\left(\begin{array}{ccc}
			1& 1 & 2\\ 
			0& 0 & 0
		\end{array}\right)
		\left(\begin{array}{ccc}
			1& L_1 & l\\ 
			0& 0 & 0
		\end{array}\right)
		\left(\begin{array}{ccc}
			1& L_2& l\\ 
			0& 0 & 0
		\end{array}\right)
		\left(\begin{array}{ccc}
			L_1& L_2 & 2\\ 
			0& 0 & 0
		\end{array}\right)
		\left\{\begin{array}{ccc}
			2& L_2 & L_1\\ 
			l& 1& 1
		\end{array}\right\}\nonumber\\
		&\left(1+\frac{D^2b^2_1}{2\pi^2}\int dkk^2j_0(kx)P(k)\right)^{-1}.
	\end{align}
	Now using the properties of spherical harmonics \cite{varshalovich1988quantum} we get,
	\begin{align}
		\label{sum_ylm_n12x}
		\sum_{\lambda}\,_2Y_{2\lambda}(\mathbf{\hat{n}_{12}})Y^{*}_{2\lambda}(\mathbf{\hat{x}})=\sqrt{\frac{5}{4\pi}}Y_{2-2}(\mathbf{\hat{x}};\mathbf{\hat{n}_{12}}).
	\end{align}
	Using Eq.(\ref{sum_ylm_n12x}) in Eq.(\ref{p_pairwise1}), we get,
	\begin{align}
		\label{p_pairwise2}		
		&P_{\mathrm{pairwise}}(\mathbf{x},\mathbf{\hat{n}_{12}}|m,\chi)=\Bigg[\frac{\sqrt{\pi}}{10\pi^5}D^4(Hfa)^2\,
		\tau_{\mathrm{eff}}\;Y_{2-2}(\mathbf{\hat{x}};\mathbf{\hat{n}_{12}})\sum_{L_1,L_2}\sum^{2}_{q=0}\sum_{l}i^{(L_1+L_2)}(-1)^{(L_1+1)}\nonumber\\
		&(2L_1+1)(2L_2+1)(2l+1)\frac{q!}{(q-l)!!(q+l+1)!!}
		\left(\begin{array}{ccc}
			1& L_1 & l\\ 
			0& 0 & 0
		\end{array}\right)
		\left(\begin{array}{ccc}
			1& L_2& l\\ 
			0& 0 & 0
		\end{array}\right)
		\left(\begin{array}{ccc}
			L_1& L_2 & 2\\ 
			0& 0 & 0
		\end{array}\right)
		\left\{\begin{array}{ccc}
			2& L_2 & L_1\\ 
			l& 1& 1
		\end{array}\right\}\nonumber\\
		&\int dk_1dk_2k_1^2k_2^2\; G_{q}(k_1,k_2,b_1,b_2)j_{L_1}(k_1x)\,j_{L_2}(k_2x)P(k_1)P(k_2)\Bigg]\Bigg[1+\frac{D^2b_1^2}{2\pi^2}\int dkk^2j_0(kx)P(k)\Bigg]^{-1}.
	\end{align}
We can rewrite the above equation in a simpler form by considering,
\begin{align}
	&C_1(x)=\sum_{L_1,L_2}\sum^{2}_{q=0}\sum_{l}\,i^{(L_1+L_2)}(-1)^{(L_1+1)}(2L_1+1)(2L_2+1)(2l+1)
	\left(\begin{array}{ccc}
		1& L_1 & l\\ 
		0& 0 & 0
	\end{array}\right)
\nonumber\\
&	\left(\begin{array}{ccc}
		1& L_2& l\\ 
		0& 0 & 0
	\end{array}\right)
	\left(\begin{array}{ccc}
		L_1& L_2 & 2\\ 
		0& 0 & 0
	\end{array}\right)
	\left\{\begin{array}{ccc}
		2& L_2 & L_1\\ 
		l& 1& 1
	\end{array}\right\}\int dk_1dk_2k_1^2k_2^2\;j_{L_1}(k_1x)\,j_{L_2}(k_2x)P(k_1)P(k_2)\nonumber\\
	&\; \frac{q!}{(q-l)!!(q+l+1)!!}G'_{q}(k_1,k_2),
\end{align}
where, 
\begin{align}
	&G'_{0}(k_1,k_2)= \frac{5}{7}\frac{1}{k_1k_2},\\
	&G'_{1}(k_1,k_2)=\frac{1}{2}\left(\frac{1}{k_1^2}+\frac{1}{k_2^2}\right),\hspace{1cm}\mathrm{and}\\
	&G'_{2}(k_1,k_2)=\frac{2}{7}\left(\frac{1}{k_1k_2}\right).
\end{align}
\begin{align}
	&C_2(x)=\frac{1}{2}\sum_{L_1,L_2}\,i^{(L_1+L_2)}(-1)^{(L_1+1)}(2L_1+1)(2L_2+1)
	\left(\begin{array}{ccc}
		1& L_1 & 0\\ 
		0& 0 & 0
	\end{array}\right)
	\left(\begin{array}{ccc}
		1& L_2& 0\\ 
		0& 0 & 0
	\end{array}\right)	\nonumber\\
&	\left(\begin{array}{ccc}
		L_1& L_2 & 2\\ 
		0& 0 & 0
	\end{array}\right)
	\left\{\begin{array}{ccc}
		2& L_2 & L_1\\ 
		0& 1& 1
	\end{array}\right\}\int dk_1dk_2k_1k_2\;j_{L_1}(k_1x)\,j_{L_2}(k_2x)P(k_1)P(k_2).
\end{align}
\begin{align}
	C_3(x)=\frac{1}{2\pi^2}\int dkk^2j_0(kx)P(k).
\end{align}
Using the above functions $C_1$, $C_2$, and $C_3$, we can rewrite Eq.(\ref{p_pairwise2}) as,
	\begin{align}
	&P_{\mathrm{pairwise}}(\mathbf{x},\mathbf{\hat{n}_{12}}|m,\chi)=A\left[(D^2Hfa)^2\right]\hspace{-0.15cm}(\chi)\,\tau_{\mathrm{eff}}(m,\chi)\nonumber\\
	&\hspace{5.7cm} \left[\frac{b_1(m,\chi)\,C_1(x)+b_2(m,\chi)\,C_2(x)}{1+D^2(\chi)b^2_1(m,\chi)\,C_3(x)}\right]Y_{2-2}(\mathbf{\hat{x}};\mathbf{\hat{n}_{12}}),
\end{align}
where $A=\frac{\sqrt{\pi}}{10\pi^5}$ is numerical constant.

	\section{Electron density profile of the galaxy clusters \label{App:cluster_profile}}
	We adopt a beta model \cite{2010ApJ_betamodel} for our cluster profile which is a spherically symmetric  model with $\beta=2/3$. It is given by,
	\begin{align}
		\label{defn_ne}
		n_e(r)=n_e(0)\left[1+\left(\frac{r}{r_c}\right)^2\right]^{-1},
	\end{align}
	where $r_c$ is the scale radius. We assumed $r_c=0.2\,R_{500c}$. The $R_{500c}$ is the  radius corresponding to the spherical overdensity mass $M_{500c}$ which we sampled using the halo mass function. The average baryon density is given by, $\rho_b= \mu_p\,m_p\, \bar{n}_e$, where $\bar{n}_e$ is volume averaged electron number density, $\mu_p=1.14$ is the average number of baryons per electron, and $m_p$ is the mass of a proton. We assume that within the radius $R_{500c}$, the ratio between the average baryon density to the average matter density follows the cosmic ratio, i.e. $\rho_m/\rho_b=\Omega_m/\Omega_b$. Therefore, from the definition of $M_{500c}$, we get,
	\begin{align}
		\label{norm_def}
		M_{500c}&=\frac{4}{3}\pi R^3_{500c}\,\rho_m\nonumber\\
		&=\left(\frac{4}{3}\pi R^3_{500c}\right)\left(\frac{\Omega_m}{\Omega_b}\right)\mu_p\,m_p \bar{n}_e\nonumber\\
		&=\left(\frac{4}{3}\pi R^3_{500c}\right)\left(\frac{\Omega_m}{\Omega_b}\right)\,1.14\,m_p \left(\frac{1}{\frac{4}{3}\pi R^3_{500c}}\right)\int_{0}^{R_{500c}}n_e(r)r^2dr\nonumber\\
		&=\frac{\Omega_m}{\Omega_b}\,1.14\,m_p\int_{0}^{R_{500c}}n_e(r)r^2dr.
	\end{align}
	Using Eq.(\ref{defn_ne}) and Eq.(\ref{norm_def}) we can fix the central electron number density $n_e(0)$. Integrating we get,
	\begin{align}
		n_e(0)=\left(\frac{1}{4\pi\cdot1.14\cdot0.03}\right)\left(\frac{\Omega_b}{\Omega_m}\right)\left(\frac{M_{500c}}{m_p}\right)\left(\frac{1}{R^3_{500c}}\right).
	\end{align}
	We should emphasize that our results are not very sensitive to the exact profile used, but only to the total mass or average optical depth of the cluster.
	\section{Estimation of cluster pairs and cluster-galaxy pairs \label{App:N_pair}}
We want to estimate the number of cluster pairs and cluster-galaxy pairs at a particular separation bin centred around $x$ with a width $\Delta x$ in a random sample. In observations, we observe clusters at higher redshifts as we look further away. For a random sample, the clusters and galaxies will be uniformly distributed in the angular space but along the radial direction, they should be distributed according to the joint distribution defined in Eq.(\ref{cl_dist}) for clusters and similarly for the galaxies with appropriate selection function for the galaxy survey. We give the derivation for cluster-cluster pairs first. Therefore, we first divide the redshift space in small bins, $\Delta z$. Corresponding to each redshift bin there is an associated comoving volume depending on the fraction of sky $\mathrm{f_{sky}}$ observed by a particular survey. We can integrate the halo number density over that redshift bin and a given mass range to find the total number of clusters in that redshift bin. 
\begin{align}
	\label{N_clusters}
	\mathrm{N_{cl}}(z)=4\pi\;\mathrm{f_{sky}} \int_{z-\Delta z}^{z+\Delta z}dz\frac{\chi^2(z)}{H(z)}\int_{M_{\mathrm{min}}}^{\infty}\frac{dn}{dm}\,dm.
\end{align}
and the cluster number density in that redshift bin is given by integrating the mass function.
\begin{align}
	n_\mathrm{cl}(z)= \int_{M_{\mathrm{min}}}^{\infty}\frac{dn}{dm}\,dm\Big|_z
\end{align} 
Then for each cluster in that redshift bin we can find the number of clusters within a shell of width $\Delta x$ at $x$ distance away assuming uniform distribution, provided $x$ is much smaller than the dimensions of the survey volume. Therefore, at each redshift bin the total number of clusters at a specific distance away from each other is given by,
\begin{align}
	\mathrm{N^{random}_{pair}}\,(x)\Big|_z\equiv RR(x)\Big|_z=\frac{1}{2}\,4\pi\, x^2\Delta x\, n_\mathrm{cl}\, \mathrm{N_{cl}}(z).
\end{align}
The factor of $1/2$ prevents double counting. Therefore, the total number of pairs in a real sample where clustering is present by using the two point correlation function $\xi(x)$ and mass-averaged linear bias factor of the clusters (halo), $\bar{b}_1$ as,
\begin{align}
	\mathrm{N^{real}_{pair}}\,(x)\Big|_z\equiv DD(x)\Big|_z=\left(1+(\bar{b}_1)^2\,\xi(x)\right)\mathrm{N^{random}_{pair}}\,(x)\Big|_z.
\end{align}
Thus, the total number of clusters at a given separation $x$ for a real sample, $DD(x)$, for cluster pairs can be obtained by summing over all the pairs at each redshift. 
\begin{align}
DD(x)=\sum_{z}	DD(x)\Big|_z=\frac{1}{2}\sum_z \left(1+(\bar{b}_1)^2\,\xi(x)\right)(4\pi\, x^2\Delta x)\, n_\mathrm{cl}(z)\, \mathrm{N_{cl}}(z).
\end{align}
Similarly we can obtain the number of cluster galaxy pair in a random sample. We need the number density of galaxies around a cluster, $n_\mathrm{gal}$. Different galaxy surveys will have their own specifications about galaxy number density depending on its observing capabilities. The total number of clusters at the redshift bin $z-\Delta z$ to $z+\Delta z$ is still given by $\ref{N_clusters}$. Therefore the number of cluster-galaxy pair the redshift bin is given by,
\begin{align}
	N^{\mathrm{random}}_\mathrm{cl-gal}(x)\Big|_z=4\pi\, x^2\Delta x\, n_\mathrm{gal}\, \mathrm{N_{cl}}(z).
\end{align}
Note that there is no factor of $1/2$ in the case of cluster-galaxy pairing, as we are calculating number of galaxies around each cluster. In this case the total number of  cluster-galaxy pairs in a real sample will depend on both linear bias factor of clusters and galaxies. Denoting the mass-averaged linear bias factor of galaxies as $\bar{b}^{\mathrm{gal}}_1$, we get,
\begin{align}
N^{\mathrm{real}}_\mathrm{cl-gal}(x)\Big|_z=\left(1+\bar{b}_1\,\bar{b}^{\mathrm{gal}}_1\,\xi(x)\right)	N^{\mathrm{random}}_\mathrm{cl-gal}(x)\Big|_z.
\end{align}
Therefore, the total number of cluster-galaxy pairs at a given separation $x$ for a real sample is,
\begin{align}
	DD(x)=\sum_{z}	N^{\mathrm{real}}_\mathrm{cl-gal}(x)\Big|_z=\sum_z \left(1+\bar{b}_1\,\bar{b}^{\mathrm{gal}}_1\,\xi(x)\right)(4\pi\, x^2\Delta x)\, n_\mathrm{gal}(z)\, \mathrm{N_{cl}}(z).
\end{align}
\color{black}In the case of LSST ``gold" survey, the redshift distribution of galaxies are given by \cite{2009_LSST}, 
\begin{align}
	p(z)=\frac{1}{2z_0}\left(\frac{z}{z_0}\right)^2\exp(-z/z_0)
\end{align}
and total number of galaxies are given by,
\begin{align}
	\mathrm{N}_{gal}=46\times10^{0.31(i-25)} \,\mathrm{galaxies\, arcmin^{-2}}.
\end{align}
We choose $i=25.3$ and the ``gold" sample covers an angular area of 20000 deg$^2$. Therefore the number density of galaxies at a certain redshift bin is given by,
\begin{align}
	n_\mathrm{gal}(z)=\mathrm{N}_{gal}(20000\times3600)\frac{P(z)}{\mathrm{Vol}},
\end{align}
where
\begin{align}
	P(z)=\int_{z-\Delta z}^{z+\Delta z}p(z)dz
\end{align}
and 
\begin{align}
\mathrm{Vol}= \frac{4\pi}{2}\int_{z-\Delta z}^{z+\Delta z}dz\frac{\chi^2(z)}{H(z)}
\end{align}
The factor of 1/2 comes from fraction of the sky LSST survey covers and we choose a $\Delta z=0.1$. \color{black}
		\section{Derivation of unpolarised pairwise linear kSZ effect\label{App:Review_kSZ}}
We can formally express the unpolarised pairwise kSZ effect as,
	\begin{align}
		\Bigg\langle\left(\frac{\Delta T_1}{T_{\mathrm{CMB}}}\right)(\mathbf{\hat{n}_1})-	\left(\frac{\Delta T_2}{T_{\mathrm{CMB}}}\right)(\mathbf{\hat{n}_1})\Bigg\rangle_2&\equiv T_{\mathrm{pairwise}}(\mathbf{x_1},\mathbf{x_2},\mathbf{\hat{n}_2},\mathbf{\hat{n}_2},m,\chi)/T_{\mathrm{CMB}}\nonumber\\
		&=\tau_{\mathrm{eff}}\,\Big\langle \mathbf{v_{1}}({\mathbf{x_1}})\cdot\mathbf{\hat{n}_1}-\mathbf{v_{2}}({\mathbf{x_2}})\cdot\mathbf{\hat{n}_2}
		\Big\rangle_2.
	\end{align}
Considering $\mathbf{\hat{n}_1}\simeq \mathbf{\hat{n}_2}\simeq(\mathbf{\hat{n}_1}+ \mathbf{\hat{n}_2})/2=\mathbf{\hat{n}_{12}}$ we get,
	\begin{align}
	T_{\mathrm{pairwise}}(\mathbf{x_1},\mathbf{x_2},\mathbf{\hat{n}_{12}},m,\chi)=\tau_{\mathrm{eff}}\,\Big\langle \mathbf{v_{1}}({\mathbf{x_1}})-\mathbf{v_{2}}({\mathbf{x_2}})	\Big\rangle_2\cdot\mathbf{\hat{n}_{12}}.
\end{align}
The expectation value $\Big\langle \mathbf{v_{1}}({\mathbf{x_1}})-\mathbf{v_{2}}({\mathbf{x_2}})\Big\rangle_2$ can be written as,
	\begin{align}
		\label{ensemble_ksz}
		\Big\langle \mathbf{v_{1}}({\mathbf{x_1}})-\mathbf{v_{2}}({\mathbf{x_2}})\Big\rangle_2 =\frac{\Big\langle \left(\mathbf{v_{1}}-\mathbf{v_{2}}\right)(1+\delta^h_1)(1+\delta^h_2)\Big\rangle}{\Big\langle (1+\delta^h_1)(1+\delta^h_2)\Big\rangle}.
	\end{align}
Let us first evaluate the numerator.
\begin{align}
\Big\langle \left(\mathbf{v_{1}}-\mathbf{v_{2}}\right)(1+\delta^h_1)(1+\delta^h_2)\Big\rangle=&\Big\langle\mathbf{v_{1}}\delta^h_2\Big\rangle-\Big\langle\mathbf{v_{2}}\delta^h_1\Big\rangle+\mathrm{higher\,order\,terms}\nonumber\\
\color {black}=\color{black}&b_1\Big\langle\mathbf{v_{1}}^{(1)}\,\delta_2^{(1)}\Big\rangle-b_1\Big\langle\mathbf{v_{2}}^{(1)}\,\delta_1^{(1)}\Big\rangle+\mathrm{higher\,order\,terms}.
\end{align}
Moving to Fourier space, we can write,
	\begin{align}
	&\delta^{(1)}(\mathbf{x},\chi)=D(\chi)\int\frac{d^3\mathbf{k}}{(2\pi)^3}\exp\left(i \mathbf{k}\cdot\mathbf{x}\right)\delta^{(1)}(\mathbf{k}),
\end{align}
\begin{align}
	&\mathbf{v}^{(1)}(\mathbf{x},\chi)=D(\chi)[afH](\chi)\int\frac{d^3\mathbf{k}}{(2\pi)^3}\exp\left(i \mathbf{k}\cdot\mathbf{x}\right) \frac{i\mathbf{k}}{k^2}\delta^{(1)}(\mathbf{k}),
\end{align}
If we make a substitution $\mathbf{k_1}\rightarrow-\mathbf{k_1}$, then $\Big\langle\mathbf{v_{2}}^{(1)}\,\delta_1^{(1)}\Big\rangle$ becomes equal to the negative of $\Big\langle\mathbf{v_{1}}^{(1)}\,\delta_2^{(1)}\Big\rangle$. Therefore, we need to evaluate only $\Big\langle\mathbf{v_{1}}^{(1)}\,\delta_2^{(1)}\Big\rangle$.
\begin{align}
	\Big\langle\mathbf{v_{1}}^{(1)}\,\delta_2^{(1)}\Big\rangle&=D^2(afH)\int\int\frac{d^3\mathbf{k_1}d^3\mathbf{k_2}}{(2\pi)^6}\left(\frac{i\mathbf{k_1}}{k_1^2}\right)\exp\left(i\mathbf{k_1}\cdot\mathbf{x_1}\right)\exp\left(i\mathbf{k_2}\cdot\mathbf{x_2}\right)\Big\langle\delta_1^{(1)}(\mathbf{k_1})\delta_1^{(1)}(\mathbf{k_2})\Big\rangle\nonumber\\
	&=D^2(afH)\int\frac{d^3\mathbf{k_1}}{(2\pi)^3}\left(\frac{i\mathbf{k_1}}{k_1^2}\right)\exp\left(i\mathbf{k_1}\cdot(\mathbf{x_1}-\mathbf{x_2})\right)P(k_1)\nonumber\\
	&=D^2(afH)\int\frac{ dk_1}{(2\pi)^3}d\Omega_{\mathbf{\hat{k}_1}}\,i\mathbf{k_1}\exp\left(i\mathbf{k_1}\cdot\mathbf{x}\right)P(k_1)\nonumber\\
	&=D^2(afH)\int\frac{ dk_1}{(2\pi)^3}\,d\Omega_{\mathbf{\hat{k}_1}}\,\frac{\partial}{\partial\mathbf{x}}\exp\left(i\mathbf{k_1}\cdot\mathbf{x}\right)P(k_1),
\end{align}
where $\mathbf{x}=\mathbf{x_1}-\mathbf{x_2}$. Expanding $\exp\left(i\mathbf{k_1}\cdot\mathbf{x}\right)$ in terms of the spherical harmonics and the Bessel function we get,
\begin{align}
	\exp(i\mathbf{k_1}\cdot\mathbf{x})=4\pi\sum_{L_1,M_1}i^{L_1}Y^{*}_{L_1M_1}(\mathbf{\hat{k}_1})Y_{L_1M_1}(\mathbf{\hat{x}})j_{L_1}(k_1x).
\end{align}
Therefore,
 \begin{align}
 \int d\Omega_{\mathbf{\hat{k}_1}}\frac{\partial}{\partial\mathbf{x}}\exp(i\mathbf{k_1}\cdot\mathbf{x})&=4\pi\sum_{L_1,M_1}i^{L_1}\int d\Omega_{\mathbf{\hat{k}_1}}Y^{*}_{L_1M_1}(\mathbf{\hat{k}_1})\frac{\partial}{\partial\mathbf{x}}\left(Y_{L_1M_1}(\mathbf{\hat{x}})j_{L_1}(k_1x)\right)\nonumber\\&=4\pi\sum_{L_1,M_1}i^{L_1}\sqrt{4\pi}\,\delta_{L_1,0}\delta_{M_1,0}\frac{\partial}{\partial\mathbf{x}}\left(Y_{L_1M_1}(\mathbf{\hat{x}})j_{L_1}(k_1x)\right)\nonumber\\
 &=4\pi\,\frac{\partial}{\partial\mathbf{x}}j_{0}(k_1x)=-4\pi\,k_1\,j_{1}(k_1x)\mathbf{\hat{x}}.
 \end{align}
Therefore,
\begin{align}
	\label{corr_vd}
	\Big\langle\mathbf{v_{1}}^{(1)}\,\delta_2^{(1)}\Big\rangle&=-D^2(afH) \frac{1}{2\pi^2}\int dk_1 k_1\,j_{1}(k_1x)P(k_1)\;\mathbf{\hat{x}}.
\end{align}
Let us now look at the volume average of the matter density two-point correlation function $\xi(x)=\big\langle\delta_1^{(1)}\delta_2^{(1)}\big\rangle$.
\begin{align}
	\xi(x)=\frac{D^2}{2\pi^2}\int dk_1 k^2_1\,j_{0}(k_1x)P(k_1).
\end{align}
Therefore, the volume average of $\xi(x)$ is,
\begin{align}
	\label{xi_bar}
	\bar{\xi}(x)&=\frac{3}{x^3}\int_{0}^{x}\xi(y)y^2dy\nonumber\\
	&=D^2\frac{1}{2\pi^2}\int dk_1 k^2_1\color{black}\,P(k_1)\color{black}\;\frac{3}{x^3}\int_{0}^{x}\,j_{0}(k_1y)y^2dy\nonumber\\
	&=D^2\frac{1}{2\pi^2}\int dk_1 k^2_1\color{black}\,P(k_1)\color{black}\;\frac{3}{x^3}\left(\frac{x^2}{k_1}j_{1}(k_1x)\right)\nonumber\\
	&=D^2\frac{1}{2\pi^2}\int dk_1 k_1\color{black}\,P(k_1)\color{black}\;\frac{3}{x}\left(j_{1}(k_1x)\right).
\end{align}
Comparing Eq.(\ref{corr_vd}) and Eq.(\ref{xi_bar}) we get,
\begin{align}
		\Big\langle\mathbf{v_{1}}^{(1)}\,\delta_2^{(1)}\Big\rangle=-\frac{1}{3}	(afH)\bar{\xi}(x)\;\mathbf{x}.
\end{align}
Therefore,
\begin{align}
	\Big\langle \left(\mathbf{v_{1}}-\mathbf{v_{2}}\right)(1+\delta^h_1)(1+\delta^h_2)\Big\rangle=-\frac{2}{3}[afH](\chi)\,b_1(\chi,m)	\bar{\xi}(x)\;\mathbf{x}.
\end{align}
Similarly we can solve the denominator in Eq.(\ref{ensemble_ksz}).
\begin{align}
	\label{deno_expansion_2}
	\Big\langle (1+\delta^h_1)(1+\delta^h_2)\Big\rangle&=1+D^2b^2_1\int\int\frac{d^3\mathbf{k_1}d^3\mathbf{k_2}}{(2\pi)^6}\exp\left(i\mathbf{k_1}\cdot\mathbf{x_1}\right)\exp\left(i\mathbf{k_2}\cdot\mathbf{x_2}\right)\Big\langle\delta_1^{(1)}(\mathbf{k_1})\delta_1^{(1)}(\mathbf{k_2})\Big\rangle\nonumber\\
	&=1+\frac{D^2b^2_1}{2\pi^2}\int dkk^2j_0(kx)P(k)=1+b^2_1\,\xi(x).
\end{align}
Therefore, the expectation value $\Big\langle \mathbf{v_{1}}({\mathbf{x_1}})-\mathbf{v_{2}}({\mathbf{x_2}})\Big\rangle_2$ becomes,
\begin{align}
\Big\langle \mathbf{v_{1}}({\mathbf{x_1}})-\mathbf{v_{2}}({\mathbf{x_2}})\Big\rangle_2=-\frac{2}{3}[afH](\chi)\frac{\,b_1(\chi,m).	\bar{\xi}(x)}{1+b^2_1\,\xi(x)}\;\mathbf{x}.
\end{align}
Finally, we can write the unpolarised pairwise kSZ signal as,
\begin{align}
	\label{ksz_final}
		T_{\mathrm{pairwise}}(\mathbf{x},\mathbf{\hat{n}_{12}},m,\chi)/T_{\mathrm{CMB}}=-\frac{2}{3}[afH](\chi)\tau_{\mathrm{eff}}(m,\chi)\frac{\,b_1(\chi,m)	\bar{\xi}(x)}{1+b^2_1\,\xi(x)}\;\left(\mathbf{x}\cdot\mathbf{\hat{n}_{12}}\right).
\end{align}
As before, Eq.(\ref{ksz_final}) is specified for a given mass and redshift of a cluster pair. We can further average over the mass and redshift (or comoving distance) distribution of cluster pairs to obtain the final signal.
	\bibliographystyle{unsrtads}
	%\bibliographystyle{unsrtnat}
	%\usepackage{babel}
	%\usepackage{biblatex}
	%\addbibresource{\references.bib}
	%	\nocite{*}
	\bibliography{references.bib}
	%\printbibliography	
	%	\begin{thebibliography}{99}
		
		%% Signal for 3 diff masses. .......2) S/N to reach 1......3)S/N for CMB-s4....4) optical depth.  ____normal way.

		% Please avoid comments such as "For a review'', "For some examples",
		% "and references therein" or move them in the text. In general,
		% please leave only references in the bibliography and move all
		% accessory text in footnotes.
		
		% Also, please have only one work for each \bibitem.

		%	\end{thebibliography}
\end{document}